\documentclass[aps,prc,10pt,superscriptaddress,twocolumn]{revtex4}
\usepackage{graphicx}
\usepackage{amssymb}
\usepackage{amsmath}
\usepackage[normalem]{ulem}
\usepackage{url}
\usepackage{color}
\usepackage{multirow}
\usepackage{xspace}
\usepackage{amsthm}
\usepackage{mathrsfs}
\usepackage[T1]{fontenc}
\usepackage{makecell}
\usepackage{hyperref}

\begin{document}

\title{Constraining the nuclear symmetry energy with electric dipole polarizability and neutron skin characteristics in \texorpdfstring{$^{208}\mathrm{Pb}$}{208Pb} within antisymmetrized molecular dynamics
}

\author{Dandan Niu}
\affiliation{China Institute of Atomic Energy, Beijing 102413, China}

\author{Xinyu Wang}
\affiliation{China Institute of Atomic Energy, Beijing 102413, China}

\author{Ying Cui}
\affiliation{China Institute of Atomic Energy, Beijing 102413, China}

\author{Qiang Zhao}
\affiliation{China Institute of Atomic Energy, Beijing 102413, China}

\author{Kai Zhao}
\affiliation{China Institute of Atomic Energy, Beijing 102413, China}

\author{Akira Ono}
\affiliation{Department of Physics, Graduate School of Science, Tohoku University, Sendai 980-8578, Japan}

\author{Yingxun Zhang}
\email{zhyx@ciae.ac.cn}
\affiliation{China Institute of Atomic Energy, Beijing 102413, China}
\affiliation{Guangxi Key Laboratory of Nuclear Physics and Technology, Guangxi Normal University, Guilin, 541004, China}
\affiliation{Southern Center for Nuclear-Science Theory (SCNT), Institute of Modern Physics, Chinese Academy of Sciences, Huizhou 516000, China}

\date{\today}

\begin{abstract}  
The electric dipole polarizability $\alpha_D$ and the neutron-skin thickness $\Delta R_{np}$ of $^{208}\mathrm{Pb}$ are two powerful and clean probes for constraining the symmetry energy at subsaturation densities. Within the framework of the antisymmetrized molecular dynamics (AMD) model, the width of the strength function and its dynamical origins are understood, and the $\alpha_D$ and $\Delta R_{np}$ data favor effective interaction parameter sets with $S_0\approx32$-34 MeV and $L=64$-87 MeV. In addition, our calculations show that the sensitive densities of $\alpha_D$ and $\Delta R_{np}$ range from 0.2$\rho_0$ to 0.57$\rho_0$, and the corresponding values of the symmetry energy at the lower and upper ends of this sensitive density region are $S(0.2\rho_0)=10.18\pm 1.10$ MeV and $S(0.57\rho_0)=22.31\pm 1.32$ MeV.
\end{abstract}

\maketitle

\section{Introduction}

The nuclear symmetry energy characterizes the isospin dependence of the isospin asymmetric nuclear equation of state, which is described by the energy per nucleon of isospin asymmetric nuclear matter. Theoretically, it can be approximately written as 
\begin{equation}
    E/A(\rho,\delta )\approx E/A(\rho,\delta =0)+S(\rho)\delta^2.
\end{equation}
The first term on the right-hand side is the equation of state for symmetric nuclear matter, and the second term is the symmetry energy, which plays a crucial role in our understanding of the properties of neutron stars, the heavy-ion collision (HIC) mechanisms and observables, and the ground-state and collective excited-state properties of nuclei. However, theoretical predictions for the density dependence of the symmetry energy $S(\rho)$ have large uncertainties away from the normal density, and thus constraining the density dependence of the symmetry energy $S(\rho)$ has become one of the main goals in nuclear physics~\cite{NuPECC2024,USDOE23LRP}.

Considerable efforts have been devoted to constraining the nuclear equation of state and nuclear symmetry energy, and comprehensive reviews on this subject can be found in Refs.~\cite{Tsang12PRC,BALi21Universe}. In nuclear astrophysics, the most commonly employed observables include the mass-radius relations and tidal deformabilities of neutron stars~\cite{BALi21Universe,Burgio21PPNP,Tsang24nature,Lattimer14EPJ,NBZhang18APJ,YXZhang20PRC}. 
In the context of HICs~\cite{Tsang09PRL,LWChen05PRL,YXZhang14PLB,BALi08PR,LWChen05PRL}, a variety of isospin-sensitive observables have been proposed to probe the symmetry energy, such as neutron-to-proton yield ratios~\cite{Famiano06PRL,JPYang21PRC}, isospin diffusion~\cite{LWChen05PRL,Tsang09PRL,ZYSun10PRC,YXZhang20PRC,Ciampi25PRC}, and charged pion yield ratios~\cite{ZGXiao09PRL,Cozma16PLB,YYLiu21PRC}. 
In nuclear structure studies, observables including the neutron-skin thickness $\Delta R_{np}$~\cite{Reed21PRL,Roca11PRL,ZZhang13PLB,Centelles09PRL,TGYue22PRR}, the isovector giant dipole resonance (IVGDR) and the electric dipole polarizability $\alpha_D$~\cite{Trippa08PRC,Roca13PRC,Colo14EPJ,ZZhang15PRC,HZheng16PRC,JXU20PRC,Reinhard21PRL}, the pygmy dipole resonance (PDR)~\cite{Klimkiewicz07PRC,Carbone10PRC,Baran12PRC,Vretenar12PRC}, and the isoscalar and isovector giant quadrupole resonances (ISGQR and IVGQR)~\cite{Roca13PRC-GQR} have been extensively used to extract constraints on the symmetry energy, particularly at subsaturation densities. For a recent review on the constraints of Nuclear equation of state from ground and collective excited state properties of nuclei, readers are referred to Ref.~\cite{ROCA18PPNP}.

Among these observables, the electric dipole polarizability $\alpha_D$ and the neutron-skin thickness $\Delta R_{np}$ are widely regarded as two particularly clean and powerful probes of the symmetry energy at subsaturation densities~\cite{Reinhard10PRC,Tsang12PRC,Roca13PRC}. A consistent description of both observables within a unified theoretical framework is therefore highly desirable. To date, two main classes of models have been employed toward this goal. One is the random-phase approximation (RPA) based on relativistic or nonrelativistic energy density functionals~\cite{Piekarewicz14EPJ,Roca15PRC,ZZhang15PRC}. Using such approaches, constraints on the symmetry energy around saturation and subsaturation densities have been extracted, for example, $S(\rho_0)=30$-35 MeV~\cite{Roca15PRC}, $S(\rho_0/3)=15.91\pm0.99$ MeV~\cite{ZZhang15PRC} and $S(\rho_0/3)=16.4\pm1.0$ MeV~\cite{JXu20PLB}. However, only a limited number of models are able to simultaneously reproduce both the IVGDR data and the PREX-II neutron skin result for $^{208}\mathrm{Pb}$, with notable exceptions given in Ref.~\cite{Reinhard21PRL}.

Another method is to use transport models, such as extended quantum molecular dynamics (EQMD)~\cite{WBHe14PRL,WBHe16PRC}, Boltzmann-Uehling-Uhlenbeck (BUU)~\cite{HYKong17PRC,JXU20PRC,RWang20PLB,YDSong23PRC}, and isospin-dependent quantum molecular dynamics (IQMD) models~\cite{CTao13PRC}. These approaches have been used to constrain the symmetry energy and are valuable for clarifying the potential reasons for the above-mentioned tension and advancing toward a definitive conclusion about its behavior in isospin asymmetric nuclear matter. 
References~\cite{JXU20PRC,HYKong17PRC,RWang20PLB,YDSong23PRC} also provide the favored constraints on the effective-mass splitting~\cite{JXU20PRC,HYKong17PRC} and in-medium nucleon-nucleon cross sections~\cite{RWang20PLB,YDSong23PRC}. 
However, it is hard to describe the neutron skin of $^{208}\mathrm{Pb}$ in QMD-like and BUU-like models, since these models are semi-classical transport models and lack strong Pauli blocking for the description of the fermion properties of nucleons as mentioned in Ref.~\cite{YXZhang18PRC}. 
The antisymmetrized molecular dynamics (AMD) model~\cite{Ono92PTP} describes the total wave function as a Slater determinant of Gaussian wave packets with fixed width, thereby preserving the fermionic character of the system and the Pauli principle. The use of such compact wave packets is advantageous for the description of fragment formation in reactions and cluster correlations in nuclei, while the full antisymmetrization provides a sound basis for the description of nuclear ground states and collective excitations. The AMD framework thus offers a unique opportunity to investigate the IVGDR and the neutron-skin thickness on the same microscopic footing within a transport model description.

In this work, we employ the antisymmetrized molecular dynamics (AMD) model~\cite{Ono92PRL,Ono92PTP} to investigate the isovector giant dipole resonance and neutron skin of $^{208}\mathrm{Pb}$. 
The paper is organized as follows. In Sec.~\ref{Sec:Theoretical framework}, we briefly describe the theoretical framework of AMD, the effective interaction parameters employed and the method for calculating the IVGDR. The main results and discussion are presented in Sec.~\ref{Sec:Results}, which includes the ground-state properties, the calculated strength functions, and the analysis of the correlations between the dipole observables and the symmetry energy. In Sec.~\ref{Sec:summary} we give a summary and outlook.

\section{Theoretical Framework }\label{Sec:Theoretical framework}

\subsection{AMD model}\label{SubSec:AMDWF}

The AMD code we used is the same as that in Ref.~\cite{Natsumi23PRC}. For convenience, we briefly introduce the framework of the AMD model and how we incorporate the perturbations and calculate the IVGDR. 

In the AMD model, the wave function of an $A$-nucleon system is given by a Slater determinant
\begin{equation}
    |\Phi\rangle=\frac{1}{\sqrt{A!}}\det[\varphi_i(j)]
\end{equation}
with
\begin{equation}
    \varphi_i=\phi_{\mathbf{Z}_i}\chi_{\alpha_i},
\end{equation}
where $\alpha_i$ represents the spin and isospin of the $i$th single-particle state, $\alpha_i=p\uparrow,p\downarrow,n\uparrow,$ or $ n\downarrow$, and $\chi$ is the spin and isospin wave function. The spatial part of the $i$th single-particle wave function $\phi_{\mathbf{Z}_i}$ is given by a localized Gaussian wave function, whose center is represented by the complex parameter $\mathbf{Z}_i$,
\begin{equation}
    \phi_{\mathbf{Z}_i}(\mathbf{r}_j)=\bigg( \frac{2\nu }{\pi}\bigg)^{3/4}\exp\bigg[ -\nu\Bigl(\mathbf{r}_j-\frac{\mathbf{Z}_i}{\sqrt{\nu}}\Bigr)^2  \bigg],
\end{equation}
where $\nu$ is a parameter to represent the width of the wave packet. For each single-particle wave function, without considering the antisymmetrization effect, it is easily seen that the real part and the imaginary part of $\mathbf{Z}_i$ correspond to the centroids of the position and the momentum, respectively,
\begin{equation}
   \mathbf{Z}_i=\sqrt{\nu}\frac{\langle \varphi_i|\mathbf{r} |\varphi_i \rangle}{\langle \varphi_i |\varphi_i \rangle} +\frac{i}{2\hbar\sqrt{\nu}}\frac{\langle \varphi_i|\mathbf{p} |\varphi_i \rangle}{\langle \varphi_i |\varphi_i \rangle}.
\end{equation}
The time  evolution of the system is determined by the time-dependent variational principle,
\begin{equation}
    \delta\int_{t_1}^{t_2}dt \frac{\langle\Phi(\mathbf{Z})|i\hbar\frac{d}{dt}-H|\Phi(\mathbf{Z})\rangle}{\langle\Phi(\mathbf{Z})|\Phi(\mathbf{Z})\rangle}=0.
\end{equation}
This leads to the equations of motion with respect to $\mathbf{Z}$,
\begin{equation}\label{Eq:EOM}
    i\hbar\sum_{j\tau} C_{i\sigma,j\tau}\dot{Z}_{j\tau}=\frac{\partial \mathcal{H}}{\partial Z_{i\sigma}^*} \quad\text{and c.c.},
\end{equation}
where $\sigma,\tau=x,y,z$ and $\mathcal{H}$ is the expectation value of the Hamiltonian $H$ with the spurious kinetic energy arising from the zero-point center-of-mass motion of the fragments subtracted~\cite{Ono92PRL,Ono92PTP},
\begin{equation}\label{Eq:Hamiltonian}
        \mathcal{H}(\mathbf{Z},\mathbf{Z}^*)=\frac{\langle\Phi(\mathbf{Z})|H|\Phi(\mathbf{Z})\rangle}{\langle\Phi(\mathbf{Z})|\Phi(\mathbf{Z})\rangle}-\frac{3\hbar^2\nu}{2M}A+T_0[A-N_F(\mathbf{Z})]
        .
\end{equation}
Here $N_F(\mathbf{Z})$ is the fragment number. In our calculations, $\nu=0.16$ fm$^{-2}$. The parameter $T_0$ is set to 8.2 MeV for the sets with $m_n^*<m_p^*$ and to 8.5 MeV for the sets with $m_n^*>m_p^*$.
The matrix $C_{i\sigma,j\tau}$ is
\begin{equation}\label{Eq:matrixC}
    C_{i\sigma,j\tau}=\frac{\partial^2}{\partial Z_{i\sigma}^*\partial Z_{j\tau}^*}\ln\langle \Phi(\mathbf{Z})|\Phi(\mathbf{Z})\rangle,
\end{equation}
which is hermitian and positive definite. The nucleon-nucleon collisions are switched off in this work. 




In the present calculations, we adopt the Skyrme-type effective interaction with the spin-orbit term omitted:
\begin{equation}
\begin{aligned}
    v_{ij}=&t_0(1+x_0P_{\sigma})\delta(\mathbf{r})\\
    &+\frac{1}{2}t_1(1+x_1P_{\sigma})[\delta(\mathbf{r})\mathbf{k}^2+\mathbf{k}^2\delta(\mathbf{r})]\\
    &+t_2(1+x_2P_{\sigma})\mathbf{k}\cdot \delta(\mathbf{r})\mathbf{k}\\
    &+\frac{1}{6}t_3(1+x_3P_{\sigma})[\rho(\mathbf{r}_i)]^{\gamma}\delta(\mathbf{r}),
\end{aligned}
\end{equation}
where $\mathbf{r}=\mathbf{r}_i-\mathbf{r}_j$ and $\mathbf{k}=\frac{1}{2\hbar}(\mathbf{p}_i-\mathbf{p}_j)$. 


To systematically investigate the influence of the symmetry energy coefficient $S_0=S(\rho_0)$ and the slope of symmetry energy $L=3\rho_0(\frac{\partial S}{\partial \rho })|_{\rho=\rho_0}$ on the IVGDR and $\Delta R_{np}$, we need to construct a set of effective interactions in which $S_0$ and $L$ can be varied with fixed incompressibility $K_0$, the isoscalar effective mass $m_s^*$, and the isovector effective mass $m_v^*$. Thus, we extend the standard Skyrme interaction by introducing an additional density-dependent interaction term~\cite{Natsumi16PRC}, i.e., 
\begin{equation}
\begin{aligned}
    v_{ij}^{\text{ext}}=\frac{1}{6}t_3x_3'P_\sigma\delta(\mathbf{r})(\rho(\mathbf{r}_i)^\gamma -\rho_0^\gamma).
\end{aligned}
\end{equation}
With this additional term, the symmetry energy takes the form
\begin{equation}
    \begin{aligned}
        S(\rho)&=\frac{1}{3}\frac{\hbar^2}{2m}\bigg(\frac{3\pi^2}{2}\bigg)^{2/3}\rho^{2/3}-\frac{1}{8}t_0(2x_0+1)\rho\\
        &\quad-\frac{1}{24}\bigg(\frac{3\pi^2}{2}\bigg)^{2/3}(3\Theta_v-2\Theta_s)\rho^{5/3}\\
        &\quad- \frac{1}{48}t_3( 2x_3 \rho_0^{\gamma}\rho+\rho^{\gamma+1} )\\
        &\quad- \frac{1}{24}t_3 x_3' \left( \rho^{\gamma+1} - \rho_0^{\gamma}\rho \right),
    \end{aligned}
\end{equation}
where $\Theta_s=3t_1+t_2(5+4x_2)$ and $\Theta_v=t_1(2+x_1)+t_2(2+x_2)$. By varying $x_3$ and $x_3'$, $S_0$ and $L$ will change, but the values of the $K_0$, $m_s^*$, and $m_v^*$ remain unchanged. 



In the present work, two groups of effective interactions are adopted, as summarized in Table~\ref{tab:parameter_sets}. One group corresponds to $m_n^*<m_p^*$ and $f_I=m/m_s^*-m/m_v^*=0.19$, whereas the other corresponds to $m_n^*>m_p^*$ and $f_I=m/m_s^*-m/m_v^*=-0.26$. This construction allows us to explore not only the effects of $S_0$ and $L$, but also the influence of the neutron-proton effective-mass splitting. The corresponding values of $x_3$ and $x_3'$ are listed in Table~\ref{tab:parameters_x3_combined} in Appendix~\ref{Appendix:ParameterSets}.

\begin{table}[htbp]
\centering
  \caption{Parameter sets used in the calculations, $S_0$ and $L$ are in MeV.}
  \label{tab:parameter_sets}
  \begin{tabular}{c c c c c c c c c} 
  \hline
  \hline
  Sets&$m_s^*/m$ & $m_v^*/m$ & $f_I$ & $S_0$& $L$ \\
  \hline
  1-15 &$0.695$ & $0.800$ & 0.19 & [30, 32, 34] & [46, 61, 75, 92, 108]  \\
  16-30  &$0.789$ & $0.653$ &$-$0.26 & [30, 32, 34] & [46, 61, 75, 92, 108] \\
  \hline
  \hline
  \end{tabular}
\end{table}

\subsection{Isovector giant dipole resonance}\label{SubSec:ivgdr}

To calculate the IVGDR response to external fields, the following procedure is employed: preparing the ground state of the nucleus, applying an external dipole perturbation at the initial moment to the ground-state nucleus, following the subsequent time evolution of the system, and then extracting the dipole strength function. 

First, we seek the ground-state configuration, i.e., wave function $\Phi_0(\{\mathbf{Z}_i\})$ of the nucleus, which is realized by using the frictional cooling method~\cite{Ono92PRL,Ono92PTP}.
Then, the collective excitation is induced by boosting the wave function $\Phi_0(\{\mathbf{Z}_i\})$ instantaneously at $t=0$ with an external perturbative field $V_{\text{ext}}(\mathbf{r},t)=\epsilon \mathcal{M} \delta(t)$ to give the initial state $\Psi(t=0^+)$ for time-dependent calculations. The parameter $\epsilon$ is the perturbation strength. The external electric dipole field operator is defined using the spherical harmonics $Y_{1\mu}$ ($\mu=\pm 1, 0$),
\begin{equation}
    \mathcal{M}(E1,\mu)=\sum_{i=1}^A e^{\text{rec}}_i r_i Y_{1\mu}(\hat{\mathbf{r}}_i),
\end{equation}
where $e^{\text{rec}}_i$ is the recoil charge ($Ne/A$ for protons and $-Ze/A$ for neutrons). This method gives the initial state $\Psi(t=0^+)$ as
\begin{equation}\label{eq:pert-wf}
    \Psi(t=0^+)=\exp[-i\epsilon \mathcal{M}(E1,\mu)/\hbar]\Phi_0, 
\end{equation}
which is the same as in Ref.~\cite{Kanada05PRC}. 
It corresponds to simply transforming the parameters $\mathbf{Z}=\{\mathbf{Z}_1,\mathbf{Z}_2,\cdots,\mathbf{Z}_A\}$ as follows:
\begin{equation}\label{eq:pert-Z}
    \mathbf{Z}_i(t=0^+)=\mathbf{Z}_i(t=0)-i\frac{\epsilon e^{\text{rec}}\sqrt{\frac{3}{4\pi}}\mathbf{e}_\mu}{2\hbar\sqrt{\nu}},
\end{equation}
where $\mathbf{e}_\mu$ denotes the unit vector corresponding to the polarization direction. 


Thirdly, the time evolution of $|\Psi(t)\rangle$ is solved by Eq. (\ref{Eq:EOM}) and the time-dependent dipole moment is obtained as
\begin{equation}\label{eq:Dt-Fourier}
    \begin{aligned}
        D_\mu(t) &\equiv \frac{\langle\Psi(t)|\mathcal{M}(E1,\mu)|\Psi(t)\rangle}{\langle\Psi(t)|\Psi(t)\rangle}\\
        &=\sqrt{\frac{3}{4\pi}}\sum_{i=1}^A e^{\rm rec}_i \mathbf{e}_\mu\cdot \frac{\rm Re \,\mathbf{Z}_i(t)}{\sqrt{\nu}}.
    \end{aligned}
\end{equation}
$D_\mu(t)$ represents the oscillation of the proton-neutron relative motion induced by $E1$ boost with $\mu=0, \pm 1$, which corresponds to the collective dipole mode of the system.

Finally, the transition strength distribution is then obtained by the following relationship, 

\begin{equation}\label{Eq:Strength function}
\begin{aligned}
    S(\omega;E1,\mu) 
    &=-\frac{1}{\pi \epsilon} \text{Im}\int_0^{\infty} dt \langle\Psi(t)|\mathcal{M}(E1,\mu)|\Psi(t)\rangle e^{i\omega t}\\
    &=-\frac{1}{\pi \epsilon} \text{Im}\int_0^{\infty} dt D_\mu(t)  e^{i\omega t}.
\end{aligned}
\end{equation}

In practical calculation, we impose the $E1$ field along the $z$ direction ($\mu=0$). 
The corresponding dipole moment is written as
\begin{equation}\label{eq:D0t}
    D_{\mu=0}(t)=\sqrt{\frac{3}{4\pi}}e\frac{NZ}{A}(R_{p,\mu=0}(t)-R_{n,\mu=0}(t)),
\end{equation}
where $R_{p,\mu=0}(t)$ and $R_{n,\mu=0}(t)$ are the $\mu=0$ components of the centers of mass of protons and neutrons, respectively. Then, the corresponding transition strength function $S(\omega;E1,\mu=0)$ can be obtained with Eq.~(\ref{Eq:Strength function}). The total strength function will be obtained with

\begin{equation}\label{eq:S_tot}
    S(\omega;E1)=3S(\omega;E1,\mu=0).
\end{equation}
Once the strength function of the IVGDR is obtained, the electric dipole polarizability $\alpha_D$ is calculated as
\begin{equation}\label{eq:alphaD}
    \alpha_D=\frac{8\pi}{9}\int \frac{S(\omega)}{\omega}d\omega.
\end{equation}
This expression is equivalent to that used in Ref.~\cite{ZZhang16PRC}. Noting that in our convention the charge factor is already included in the definition of $S(\omega)$, whereas it is kept outside the strength function in Ref.~\cite{ZZhang16PRC}.








\section{Results and Discussions}\label{Sec:Results}

\subsection{Fluctuations in the initial configuration}

The ground state of the nucleus is prepared via the frictional cooling method as mentioned before. The cooling calculation is repeated 200 times for each interaction set. Depending on the initial conditions for the cooling calculation, different energy-minimum configurations are obtained. One possible interpretation is that the true ground-state wave function contains these configurations; in that case, it is reasonable to average the results over different configurations, as in the typical case of a well-deformed nucleus. On the other hand, if some cooling calculations are trapped in local minima that do not correspond to the ground state, such configurations should in principle be excluded. Otherwise, the resulting configuration dependence should be regarded as reflecting model uncertainties. The actual situation may be a mixture of these two extremes. For a conservative estimate, the fluctuations arising from different configurations may be included in the uncertainties of the results.

\subsection{Binding energy, charge radius, and neutron-skin thickness}\label{SubSec:Properties of GS}

Before addressing the IVGDR, it is essential to check the charge radius $r_{ch}$ and the neutron-skin thickness $\Delta R_{np}$ of $^{208}$Pb obtained with AMD. 



\begin{figure}[htbp]
	\centering
	\includegraphics[width=1.0\linewidth]{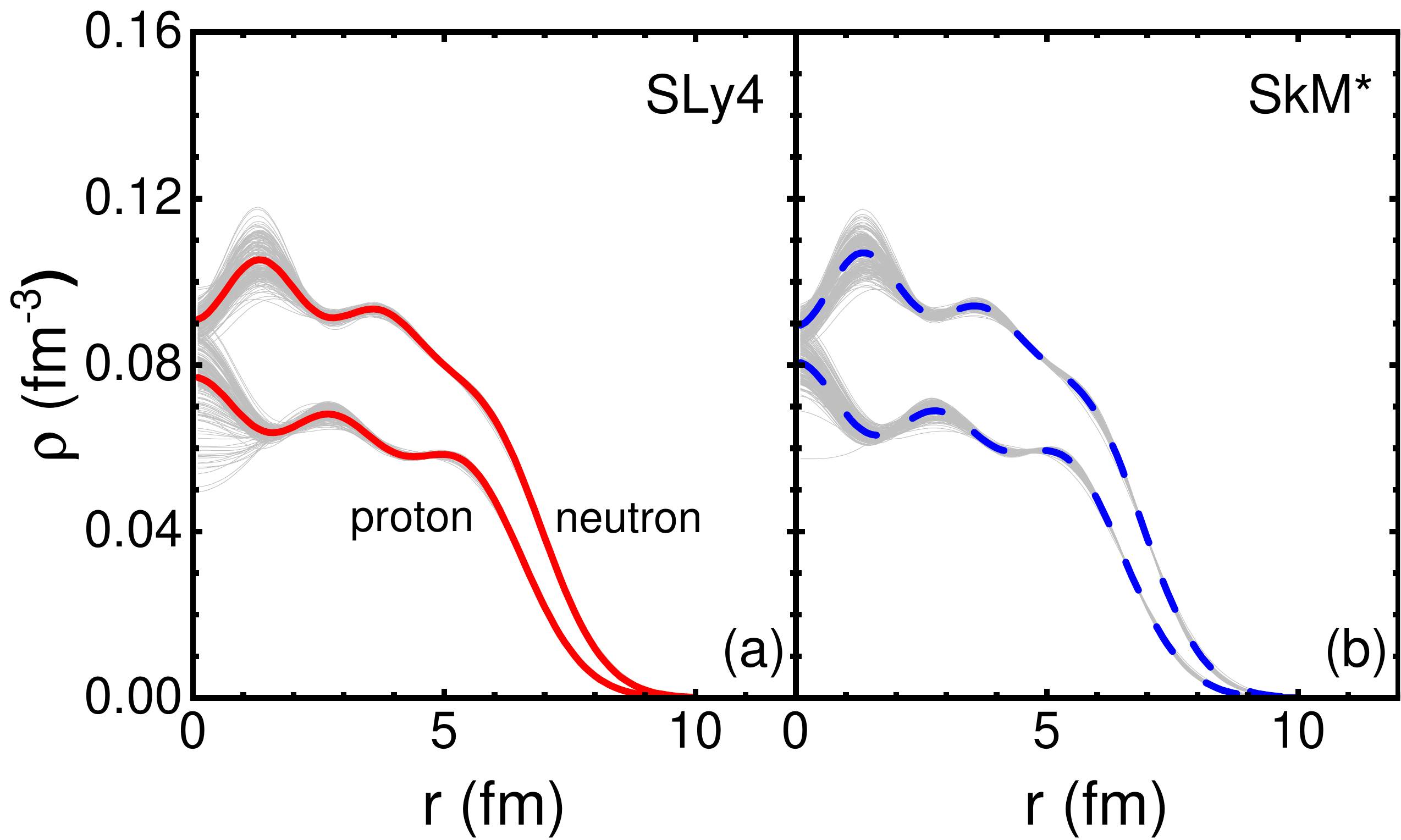}
	\caption{Radial density distributions of neutrons and protons for \texorpdfstring{$^{208}\text{Pb}$}{208Pb}. The thin lines are the results for 200 different configurations, and the thick lines are the results averaged over the configurations.}
    \label{fig:rhor-200}
\end{figure}

In practical calculations, the calculated energy and density distributions of the initial ground state $|\Phi_0\rangle$ depend on the configurations obtained by the frictional cooling method.
As an example, we plot the density profiles of $^{208}\mathrm{Pb}$ obtained with SLy4 and SkM* interactions in Fig.~\ref{fig:rhor-200}. The thin lines are the results for 200 different configurations, and the thick lines are the results averaged over the configurations.

Using the density distributions obtained for each configuration, the root-mean-square radii of protons and neutrons are calculated by removing the center-of-mass contribution,
\begin{equation} 
    \begin{aligned} 
        \langle r_p^2\rangle &= \left\langle \frac{1}{Z} \sum_{i\in p} (\mathbf{r}_i-\mathbf{R}_{\rm cm})^2 \right\rangle =\frac{1}{Z}\left\langle \sum_{i\in p}\mathbf{r}_i^2 \right\rangle-\bigg\langle \mathbf{R}_{\rm cm}^2 \bigg\rangle ,\\
        \langle r_n^2\rangle &= \left\langle \frac{1}{N} \sum_{i\in n} (\mathbf{r}_i-\mathbf{R}_{\rm cm})^2 \right\rangle =\frac{1}{N}\left\langle \sum_{i\in n}\mathbf{r}_i^2 \right\rangle-\bigg\langle \mathbf{R}_{\rm cm}^2 \bigg\rangle , 
    \end{aligned} 
\end{equation}
where $\mathbf{R}_{\rm cm}=\frac{1}{A}\sum_{i=1}^A\mathbf{r}_i$ is the center-of-mass coordinate of the nucleus, and then the charge radius and neutron-skin thickness can be obtained as follows:
\begin{equation}
\begin{aligned}
    r_{ch}&=\sqrt{\langle r_p^2\rangle +(0.8\ \text{fm})^2},\\
    \Delta R_{np}&=\langle r_n^2\rangle^{1/2}-\langle r_p^2\rangle^{1/2}.
\end{aligned}
\end{equation}
One can expect that the $E_B/A$, $r_{ch}$, and $\Delta R_{np}$ differ across configurations, corresponding to different density distributions. 
Table~\ref{tab:Pb208_ground_state} summarizes the values of $\langle E_B/A\rangle$, $\langle r_{ch}\rangle$, and  $\langle \Delta R_{np}\rangle$ for $^{208}\mathrm{Pb}$, calculated with  the 30 parameter sets listed in Table~\ref{tab:parameter_sets}.  The notation $\langle \cdot \rangle$ denotes the ensemble average over 200 configurations.
The standard deviations of these observables depend on the effective interaction parameter sets. In general, they satisfy $\delta  E_B/A <0.01$ MeV, $\delta r_{ch}<0.02$ fm, and $\delta \Delta R_{np}<0.01$ fm, which are negligibly small and can be ignored for the purposes of the present paper.

\begin{figure}[htbp]
	\centering
	\includegraphics[width=1.0\linewidth]{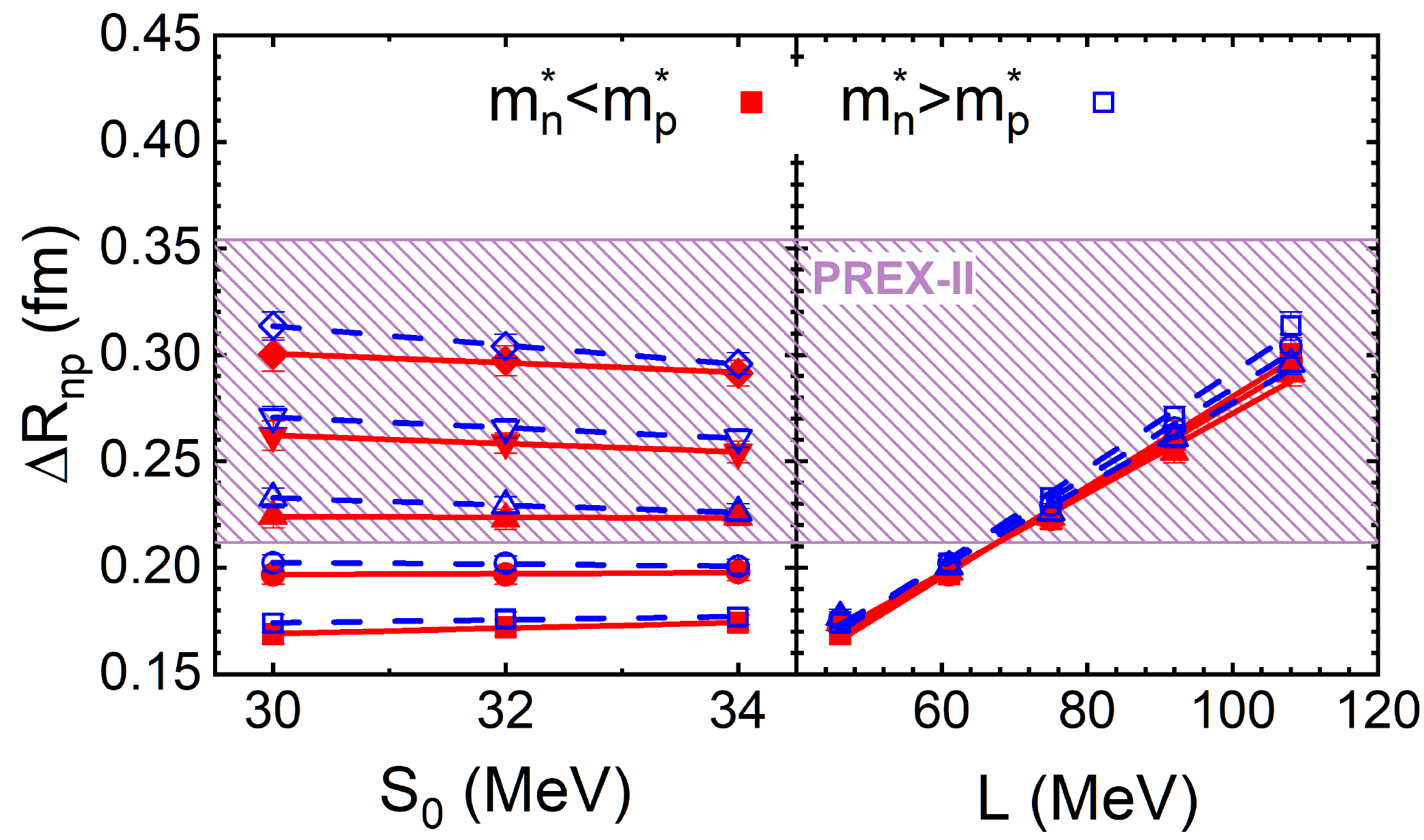}
	\caption{(a) The $\Delta R_{np}$ of $^{208}\mathrm{Pb}$ as functions of $S_0$, the lines from bottom to top correspond to L=46 to 108 MeV, and (b) $\Delta R_{np}$ as functions of $L$, the lines from bottom to top correspond to $S_0$=34 to 30 MeV. Red symbols denote parameter sets with $m_n^*<m_p^*$, while blue symbols denote those with $m_n^*>m_p^*$.}
    \label{fig:Rnp-S-L}
\end{figure}

Figures~\ref{fig:Rnp-S-L}(a) and (b) show the dependence of $\langle \Delta R_{np}\rangle $ on $S_0$ and $L$. 
For a given value of $L$, $\langle \Delta R_{np}\rangle $ weakly depends on $S_0$ and $\Delta m_{np}^*/\delta=(m_n^*-m_p^*)/m$. In contrast, for a fixed $S_0$, $\Delta R_{np}$ is strongly correlated with $L$, and $\langle \Delta R_{np}\rangle$ increases monotonically with the slope parameter $L$. This behavior is consistent with the well-known correlation that a stiffer symmetry energy leads to larger neutron skin. By comparing the calculated $\langle \Delta R_{np}\rangle $ with the PREX-II neutron-skin data~\cite{Adhikari21PRL}, we found that only the sets with $L>67$ MeV can reproduce the data.

\begin{table}[htbp]
  \centering
  \footnotesize
  \caption{The properties of \texorpdfstring{$^{208}$Pb}{208Pb} in the AMD initial ground state for all 30 parameter sets. The experimental data are listed in the last row. \texorpdfstring{$S_0$}{S0}, \texorpdfstring{$L$}{L}, and \texorpdfstring{$E_B/A$}{EB} are in MeV; \texorpdfstring{$r_{ch}$}{rch} and \texorpdfstring{$\Delta R_{np}$}{Rnp} are in fm. The symbols \texorpdfstring{$\dagger$}{dagger} and \texorpdfstring{$\ddagger$}{ddagger} denote the SLy4 and SkM* interactions, respectively.}
  \label{tab:Pb208_ground_state}
  \setlength{\tabcolsep}{4pt}
  \renewcommand{\arraystretch}{1.0}
  \begin{tabular}{ c c c c c c c}
    \hline
    \hline
     Sets &  $\Delta m_{np}^*/\delta$ ($f_I$) & $S_0$ & $L$ & $\langle E_B/A\rangle $ & $\langle r_{\text{ch}}\rangle$ & $\langle \Delta R_{np}\rangle$ \\
    \hline
   1 &  &  & 46  & 7.868 & 5.570 & 0.170 \\
   2 &  &  & 61  & 7.868 & 5.559 & $0.199$ \\
   3  &  $-$0.15 (0.19)  & 30 & 75  & 7.868 & 5.548 & $0.227$ \\
   4  &  &    & 92  & $7.869$ & 5.534 & $0.265$ \\
   5  &  &    & 108 & $7.869$ & $5.521$ & $0.301$ \\
    \hline
    6$\dagger$ &  &    & 46  & $7.857$ & $5.576$ & $0.172$ \\
    7 &  &    & 61  & 7.868 & $5.561$ & $0.198$ \\
    8 &   $-$0.15 (0.19)  & 32 & 75  & 7.868 & $5.549$ & $0.225$ \\
    9 &  &    & 92  & 7.868 & $5.536$ & $0.262$\\
    10&  &    & 108 & 7.869 & $5.520$ & $0.299$\\
    \hline
    11 &  &    & 46  & 7.818 & $5.584$ & $0.174$ \\
    12 &  &    & 61  & 7.843 & $5.568$ & $0.199$ \\
    13 &   $-$0.15 (0.19)  & 34 & 75  & 7.866 & $5.556$ & $0.223$ \\
    14 &  &    & 92  & 7.868 & $5.539$ & $0.256$ \\
    15 &  &    & 108 & 7.868 & $5.525$ & $0.292$ \\
    \hline
    \hline
    16$\ddagger$ &  &    & 46  & $7.735$ & $5.547$ & $0.174$ \\
    17 &  &    & 61  & $7.766$ & $5.531$ & $0.202$ \\
    18 &   0.30 ($-$0.26) & 30 & 75  & $7.801$ & $5.518$ & $0.232$ \\
    19 &  &  & 92  & 7.848 & $5.494$ & $0.271$ \\
    20 &  &  & 108 & 7.868 & $5.472$ & $0.316$ \\
    \hline
    21 &  &    & 46  & 7.700 & $5.552$ & $0.176$ \\
    22 &  &    & 61  & 7.728 & $5.537$ & $0.202$ \\
    23 &  0.30 ($-$0.26) & 32 & 75  & 7.758 & $5.521$ & $0.230$ \\
    24 &  &    & 92  & 7.802 & $5.502$ & $0.266$ \\
    25 &  &    & 108 & 7.852 & $5.477$ & $0.305$ \\
    \hline
    26 &  &    & 46  & 7.671 & $5.559$ & $0.177$ \\
    27 &  &    & 61  & 7.693 & $5.544$ & $0.201$ \\
    28 &  0.30 ($-$0.26) & 34 & 75  & 7.720 & $5.529$ & $0.226$ \\
    29 &  &    & 92  & $7.761$ & $5.505$ & $0.261$ \\
    30 &  &    & 108 & $7.805$ & $5.486$ & $0.296$ \\
    \hline
    \renewcommand{\arraystretch}{1.8} 
    Exp. & & & & 7.867 & 
    \makecell{$5.501$ \\ $\pm 0.001$} & 
    \makecell{$0.283$ \\ $\pm 0.071$\cite{Reed21PRL}} \\
    \hline
    \hline
  \end{tabular}
\end{table}

\subsection{Time evolution of dipole moments and the transition strength function}\label{SubSec:Calculate of IVGDR}

To calculate the IVGDR within the AMD framework, we analyze the time evolution of the dipole moments of $^{208}$Pb following an instantaneous electric dipole perturbation applied along the $z$ direction as described in Sec~\ref{SubSec:ivgdr}. 

\begin{figure}[htbp]
	\centering
	\includegraphics[width=1\linewidth]{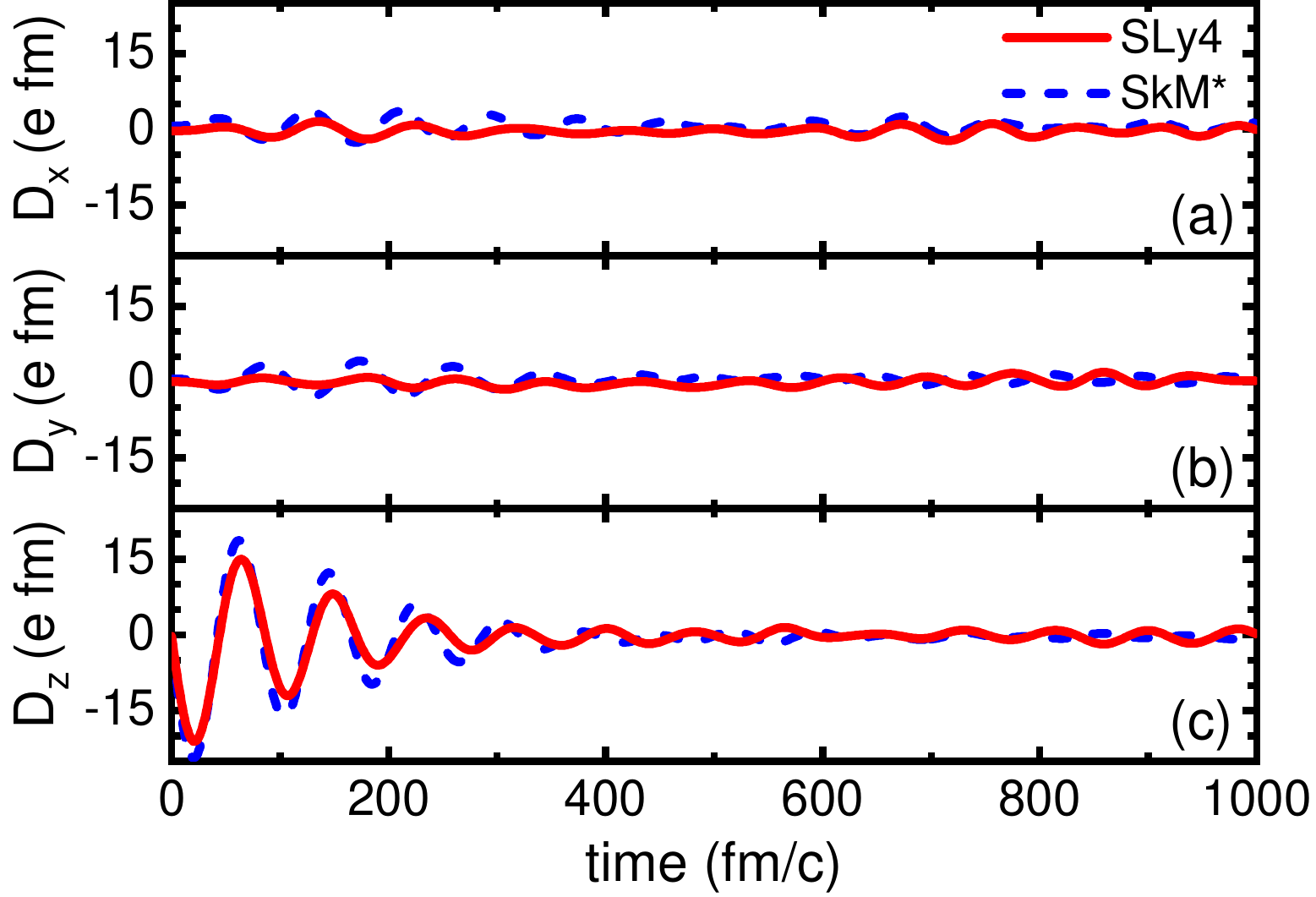}
	\caption{The time evolution of the electric dipole moment in x, y, z directions for \texorpdfstring{$^{208}$Pb}{208Pb}. Red lines are for SLy4, and blue lines are for SkM*.}
    \label{fig:Dt}
\end{figure}

In the present work, the perturbation strength is chosen as $\epsilon=50\  \text{MeV}\,c^{-1}e^{-1}$. We have verified that the oscillation frequency of the induced dipole motion is independent of $\epsilon$ for $\epsilon \lesssim 72\ \text{MeV}\,c^{-1}e^{-1}$, ensuring that all calculations are performed within the linear response regime. Figure~\ref{fig:Dt} shows the time evolution of the dipole moment components $D_x(t)$, $D_y(t)$, and $D_z(t)$ for a representative initial configuration, where
\begin{equation}
    D_i(t)=e\frac{NZ}{A}\big[R_{p,i}(t)-R_{n,i}(t)\big], \qquad i=x,y,z.
\end{equation}
The solid lines correspond to results obtained with the SLy4 interaction, while the dashed lines show those calculated with the SkM* interaction.

As shown in Figs.~\ref{fig:Dt}(a) and (b), only weak fluctuations around zero are observed in both $D_x(t)$ and $D_y(t)$. These small-amplitude components originate from the intrinsic inhomogeneity of the initial ground-state density distributions.  
In contrast, a pronounced collective oscillation is clearly seen in the $z$ component $D_z(t)$ [Fig.~\ref{fig:Dt}(c)], which reflects the isovector dipole mode induced by the external perturbation. This oscillation is driven by the restoring force associated with the symmetry potential, which is mainly localized in the nuclear surface region. 

The feature can be understood in a simple dynamical picture within the transport model, where the equation of motion of $\mathbf{D}$ can be written as
\begin{equation}
    \begin{aligned}
           \ddot{\mathbf{D}}  & =e\frac{NZ}{Am} \nabla\bigg((\frac{1}{m_s^*}-\frac{1}{m_v^*})\hbar^2k_F^2\bigg)\frac{1}{2}\frac{\nabla\rho}{\rho}\cdot \mathbf{D}\\
           &\quad +e\frac{NZ}{Am} \bigg[ \frac{2}{3}L(\rho) -4S(\rho)\bigg] \frac{(\nabla\rho)^2}{\rho^2} \mathbf{D}.\\
    \end{aligned}
\end{equation}
The derivation of the above formula can be found in Appendix~\ref{app:ddotD}. Thus, the oscillation frequency of the dipole mode can also be estimated from the force constant $k$ by approximating the restoring force as $F\simeq -kD$. This means the oscillation frequency is proportional to the values of 
\begin{equation}\label{eq:freqs}
    e\frac{NZ}{Am} \bigg[4S(\rho)-\frac{2}{3}L(\rho)\bigg] \frac{(\nabla\rho)^2}{\rho^2},
\end{equation}
and
\begin{equation}\label{eq:freq-fi}
    -e\frac{NZ}{Am} \bigg[\nabla\bigg((\frac{1}{m_s^*}-\frac{1}{m_v^*})\hbar^2k_F^2\frac{1}{2}\frac{\nabla\rho}{\rho}\bigg)\bigg]^T
\end{equation}
at the surface of the nucleus as the gradient of the single particle potential approaches zero in the interior of the nucleus. The superscript $T$ in Eq.(\ref{eq:freq-fi}) denotes the transpose of matrix in bracket $[\cdot]$. Formally, the frequency depends on the isovector effective mass $m_v^*(\rho)$, isoscalar effective mass $m_s^*(\rho)$, the slope of the symmetry energy $L(\rho)$ and the strength of the symmetry energy $S(\rho)$. 

According to Eq.~(\ref{eq:freqs}), a larger value of $S_0$ yields a larger symmetry energy over the entire density range compared with that for a smaller $S_0$, which implies a stronger restoring force and thus a higher oscillation frequency (or equivalently, a higher centroid energy) for larger $S_0$. However, the centroid energy for a larger $L$ is smaller than that for a smaller $L$. This is because a stiffer symmetry energy corresponds to a smaller symmetry energy at subsaturation densities compared with a softer symmetry energy. Consequently, a larger $L$ leads to a smaller centroid energy. The difference between the isoscalar effective mass and isovector effective mass, i.e.,  $f_I=m/m_s^*-m/m_v^*$, also affects the main frequency of the strength function. The positive value of $f_I$ will decrease the frequency, whereas the negative value of $f_I$ will increase the frequency.

As shown in Fig.~\ref{fig:Dt}, the amplitude of the oscillation gradually decreases over time due to the Landau damping mechanism. This mechanism means that the different local mean fields and effective masses experienced by individual Gaussian wave packets lead to different oscillation frequencies, and this phase-mixing mechanism leads to a gradual loss of phase coherence. This picture appears different from that in the RPA framework, where Landau damping originates from the fragmentation of the collective strength over particle-hole configurations in energy space. Thus, using the same effective interaction in AMD model does not necessarily guarantee an identical result as in RPA, because the two approaches treat the underlying many-body dynamics and the damping mechanisms in different ways.

\begin{figure}[htbp]
	\centering
	\includegraphics[width=1\linewidth]{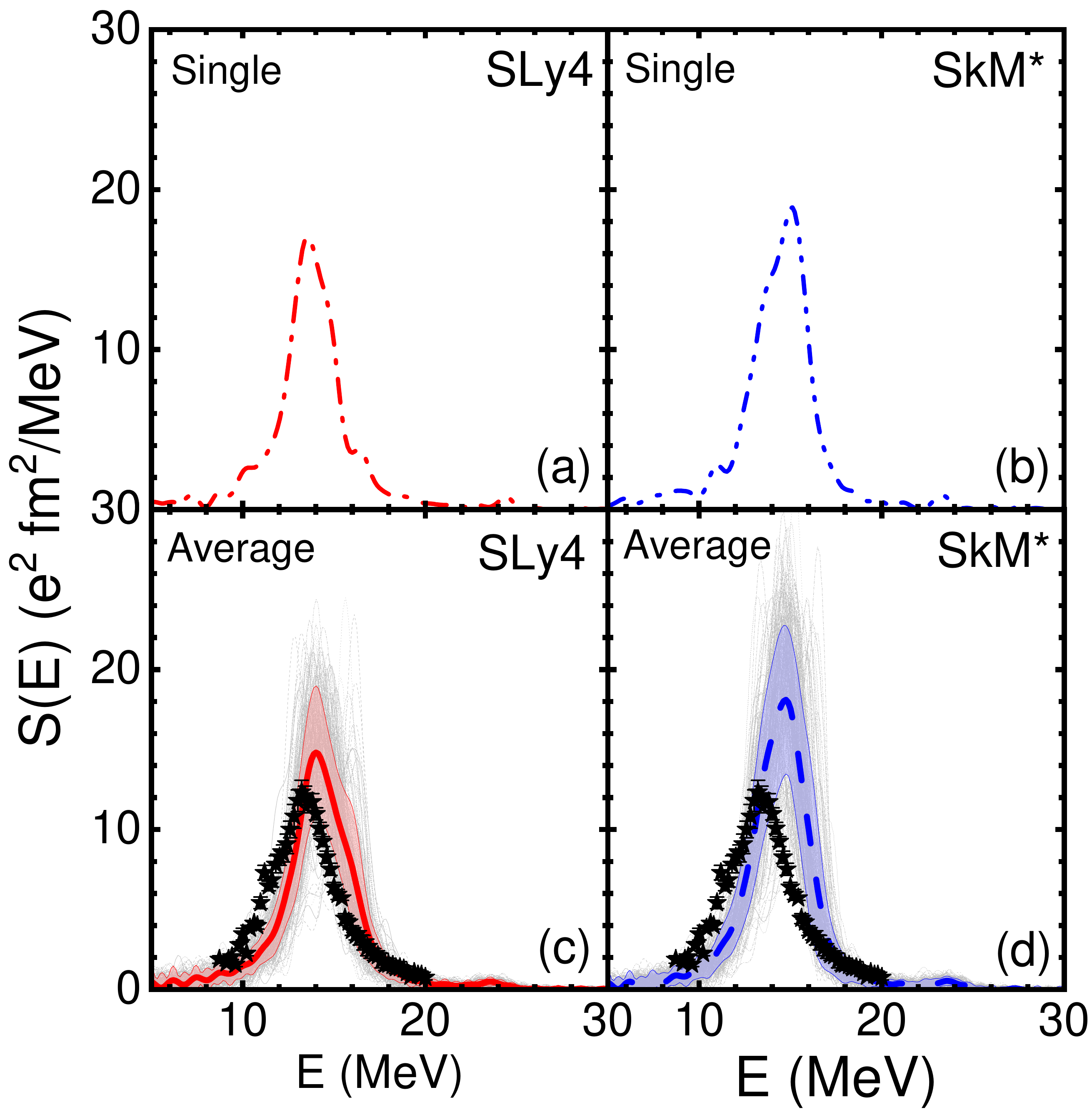}
	\caption{Isovector giant dipole resonance strength function for \texorpdfstring{$^{208}$Pb}{208Pb}. }
    \label{fig:S(E)}
\end{figure}

By using Eqs.~(\ref{eq:Dt-Fourier})$-$(\ref{eq:S_tot}), the strength function $S(E)$ can be obtained for each initial configuration. Figures~\ref{fig:S(E)}(a) and (b) are obtained from one initial configuration for SLy4 and SkM*, respectively. It clearly indicates that the dynamics generate the spreading of the strength function even for a single initial configuration. 
Figures~\ref{fig:S(E)}(c) and (d) show the results form the 200 initial configurations, where the thin colored lines correspond to individual configurations and the thick lines represent the averaged strength functions.
To quantify the variation among different initial-configuration dependence, the standard deviation of $S(E)$ is also calculated and shown as the shaded region. The averaged $S(E)$ with SLy4 effective interaction is close to the data (black points) without introducing the smoothing parameter $\Gamma$ as in Ref.~\cite{Kanada05PRC} owing to the Landau damping. although adding a width of 1 MeV can reduce the height of $S(E)$ which improves the agreement with the data. 
The above mechanism is different from the discussions in the BUU transport models in Refs.~\cite{HYKong17PRC,JXU20PRC,RWang20PLB,YDSong23PRC}, in which two-body collisions also contribute to the spreading of the strength function. The spurious two-body collisions are caused by the underestimations of the Pauli blocking probability in the BUU type models~\cite{YXZhang18PRC}. In addition, we also turn on the nucleon-nucleon collisions in the AMD model to check its influence on $S(E)$, and our calculations show that the difference in the integral of $S(E)$ is less than 0.47\% due to the well-described Pauli blocking in AMD model. 

\subsection{Influence of \texorpdfstring{$S_0$, $L$ and $f_I$ on $S(E)$}{S0, L and fI on S(E)}}\label{SubSec:Constraints on symmetry energy}

Before constraining the symmetry energy via the IVGDR, we first examine the strength function $S(E)$ obtained with two typical effective interactions, i.e., SLy4 and SkM*. Both interactions have the same $L$, but different $S_0$ values and isoscalar and isoevector effective masses (or different $f_I$ values)~\cite{Chabanat97NPA,YXZhang14PLB}. Figure~\ref{fig:S(E)}(c) shows the results obtained with the SLy4 interaction, while panel (d) corresponds to SkM*. Our calculations show that the centroid energy of $S(E)$ obtained with the SLy4 interaction ($f_I=0.19$, i.e., $m_n^*<m_p^*$) is smaller than that obtained with the SkM* interaction ($f_I=-0.26$, i.e., $m_n^*>m_p^*$). The centroid energy of $S(E)$ obtained with SLy4 is about 14.0 MeV, which is lower than that obtained with SkM* (about 14.7 MeV). This difference mainly comes from the impact of $f_I$.  

\begin{figure}[htbp]
	\centering
	\includegraphics[width=1.0\linewidth]{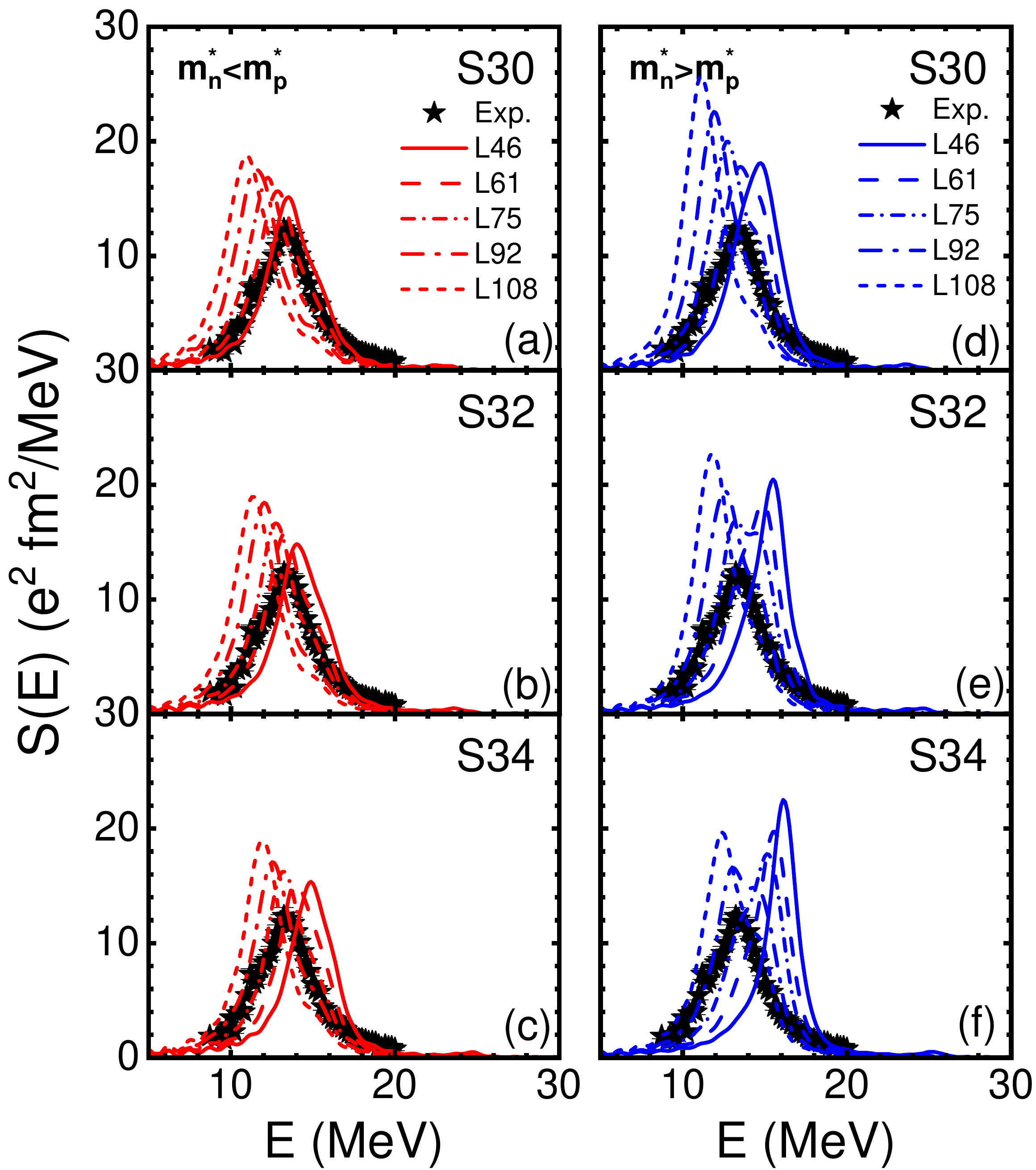}
	\caption{Calculated IVGDR strength functions and data for \texorpdfstring{$^{208}$Pb}{208Pb}.}
    \label{fig:strength-diff-L}
\end{figure}

To illustrate the influence of $S_0$, $L$, and $f_I$ on $S(E)$, we performed calculations with 30 parameter sets featuring different values of $f_I$, $S_0$, and $L$, as summarized in Table~\ref{tab:parameter_sets}, and the results are presented in Fig.~\ref{fig:strength-diff-L}. The left panels in Fig.~\ref{fig:strength-diff-L} show the $S(E)$ for sets 1$-$15 with $f_I=0.19$ ($m_n^*<m_p^*$), while the right panels correspond to the results for sets 16$-$30 with $f_I=-0.26$ ($m_n^* > m_p^*$). 
The panels from top to bottom correspond to $S_0 = 30$, 32, and 34 MeV, respectively. The lines in each panel represent the results obtained with different values of $L$. 
Our calculations clearly demonstrate that the centroid energies increase with increasing $S_0$ and decrease with increasing $L$, with all other parameters fixed. In addition, the centroid energy is systematically larger for $f_I<0 (\Delta m_{np}^* > 0)$ than for $f_I>0 (\Delta m_{np}^* < 0)$ at a given $S_0$ and $L$.

\begin{table}[htbp]
  \centering
  \footnotesize
  \caption{The calculated \texorpdfstring{$\chi_r^2$}{chi2} values for all 30 parameter sets compared with the experimental data. The symbols \texorpdfstring{$\dagger$}{dagger} and \texorpdfstring{$\ddagger$}{ddagger} denote the SLy4 and SkM* interactions, respectively. }
  \label{tab:chi2_results}
  \setlength{\tabcolsep}{8pt}
  \renewcommand{\arraystretch}{1.0}
  \begin{tabular}{c c c c c }
    \hline
    \hline
    Sets & $\Delta m_{np}^*/\delta$ ($f_I$) & $S_0$ (MeV) & $L$ (MeV) & $\chi_r^2$ \\
    \hline
    1 &   & &46  & 0.74 \\
    2 &   & &61  & 0.99 \\
    3 &$-$0.15 (0.19)  & 30 & 75  & 2.19 \\
    4 &   & &92  & 4.71 \\
    5 &   & &108 & 7.67 \\
    \hline
    6$\dagger$ &   & &46  & 2.28 \\
    7 &   & &61  & 0.94 \\
    8 &$-$0.15 (0.19) & 32 & 75  & 1.27 \\
    9 &   & &92  & 3.64 \\
    10 &   & &108 & 6.72 \\
    \hline
    11 &   & &46  & 7.44 \\
    12 &   & &61  & 2.60 \\
    13 & $-$0.15 (0.19) & 34 & 75  & 1.11 \\
    14 &   & &92  & 1.71 \\
    15 &   & &108 & 4.32 \\
    \hline
    \hline
    16$\ddagger$ &   & &46  & 3.21 \\
    17 &   & &61  & 1.31 \\
    18 &0.30 ($-$0.26) & 30 & 75  & 1.67 \\
    19 &   & &92  & 3.69 \\
    20 &   & &108 & 8.11 \\
    \hline
    21 &   & &46  & 8.52 \\
    22 &   & &61  & 3.10 \\
    23 &0.30 ($-$0.26) & 32 & 75  & 1.50 \\
    24 &   & &92  & 2.29 \\
    25 &   & &108 & 4.38 \\
    \hline
    26 &   & &46  & 19.38 \\
    27 &   & &61  & 7.12 \\
    28 & 0.30 ($-$0.26) & 34 & 75  & 3.11 \\
    29 &   & &92  & 1.52 \\
    30 &   & &108 & 2.35 \\
    \hline
    \hline
  \end{tabular}
\end{table}


To quantitatively assess the agreement between the theoretical calculations and the experimental data, we performed a $\chi^2$ analysis for all 30 parameter sets. 
The reduced $\chi_r^2$ is defined as
\begin{equation}
    \chi_r^2 = \frac{1}{N_{\rm exp}}\sum_{i=1}^{N_{\rm exp}}\frac{\left[S_{\rm th}(E_i)-S_{\rm exp}(E_i)\right]^2}{\sigma^2_{\rm th}(E_i)+\sigma_{\rm exp}^2(E_i)},
\end{equation}
where $S_{\mathrm{th}}(E_i)$ and $S_{\mathrm{exp}}(E_i)$ denote the theoretical and experimental $E1$ strength functions at energy $E_i$, respectively, $\sigma_{\rm exp}(E_i)$ is the experimental uncertainty and $\sigma_{\rm th}(E_i)$ represents standard deviation of the initial-configuration dependence at energy $E_i$. 
The detailed results are summarized in Table~\ref{tab:chi2_results}. Our calculations show that the parameter set with $f_I=0.19$, $S_0=30$ MeV, and $L=46$ MeV gives the smallest reduced chi-square value, \(\chi_r^2=0.74\). 

\subsection{Constraining symmetry energy via \texorpdfstring{$\alpha_D$}{alphaD} and \texorpdfstring{$\Delta R_{np}$}{Rnp}}

To give the constraint of symmetry energy at subsaturation densities, we perform the combination analysis on $\alpha_D$ and $\Delta R_{np}$ within the framework of the AMD model. 

\begin{figure}[htbp]
	\centering
	\includegraphics[width=1\linewidth]{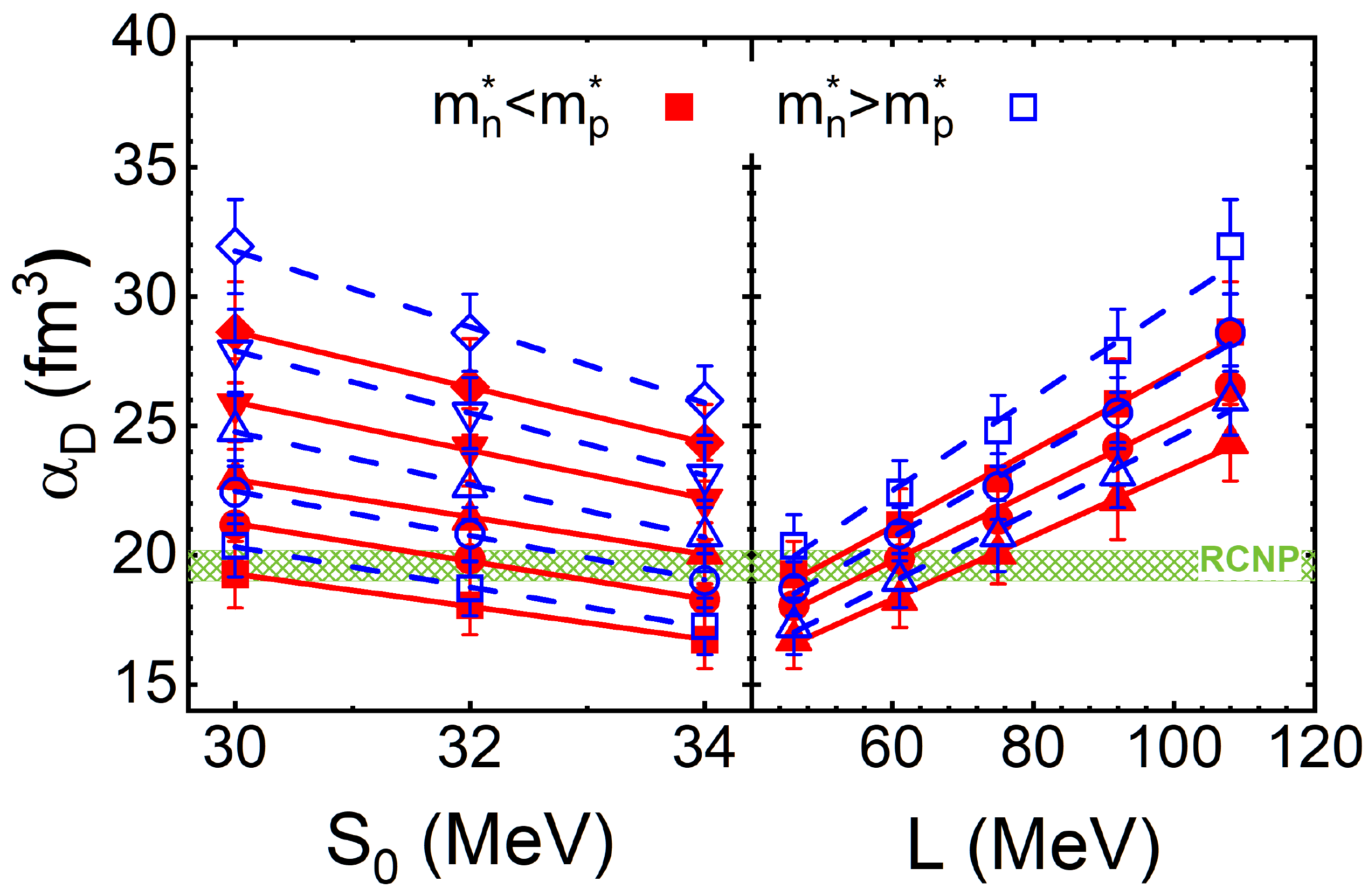}
	\caption{ Same as Fig.~\ref{fig:Rnp-S-L}, but for \texorpdfstring{$\alpha_D$}{alphaD}. The error bars denote the corresponding standard deviation over 200 configurations.}
    \label{fig:Rnp-alphaD-L}
\end{figure}

Figures~\ref{fig:Rnp-alphaD-L}(a) and (b) display the electric dipole polarizability $\alpha_D$ as a function of $S_0$ and $L$, respectively. More specifically, $\alpha_D$ for each configuration is obtained by integrating Eq.(\ref{eq:alphaD}) from the lower limit $\hbar\omega_{\text{min}} =5\rm \,MeV$, which is the same as in experiment~\cite{Tamii11PRL}, to the upper limit $\hbar\omega_{\text{max}} =60\rm \,MeV$ which guarantees that the values of $\alpha_D$ reach an asymptotic value. The symbols indicate the average value $\alpha_D$ and the error bars denote the corresponding standard deviation over 200 configurations.  
Similarly to the neutron skin, $\alpha_D$ exhibits a robust linear dependence on $L$. However, $\alpha_D$ also shows a pronounced dependence on $S_0$ and $\Delta m_{np}^*$. Specifically, for a fixed $L$, the sets with $\Delta m_{np}^*>0$ (blue lines) yields systematically higher $\alpha_D$ values than the sets with $\Delta m_{np}^*<0$ (red lines), consistent with the results reported in Ref.~\cite{HYKong17PRC}. By comparing the calculations to the $\alpha_D$ data~\cite{Tamii11PRL,Roca15PRC}, we found that the sets with $S_0=[30,34]$ and $L=[35, 86]$ are compatible with the data within the combined experimental and theoretical uncertainties. Moreover, the extracted $S_0$ is positively correlated with $L$. 

\begin{figure}[htbp]
	\centering
	\includegraphics[width=0.8\linewidth]{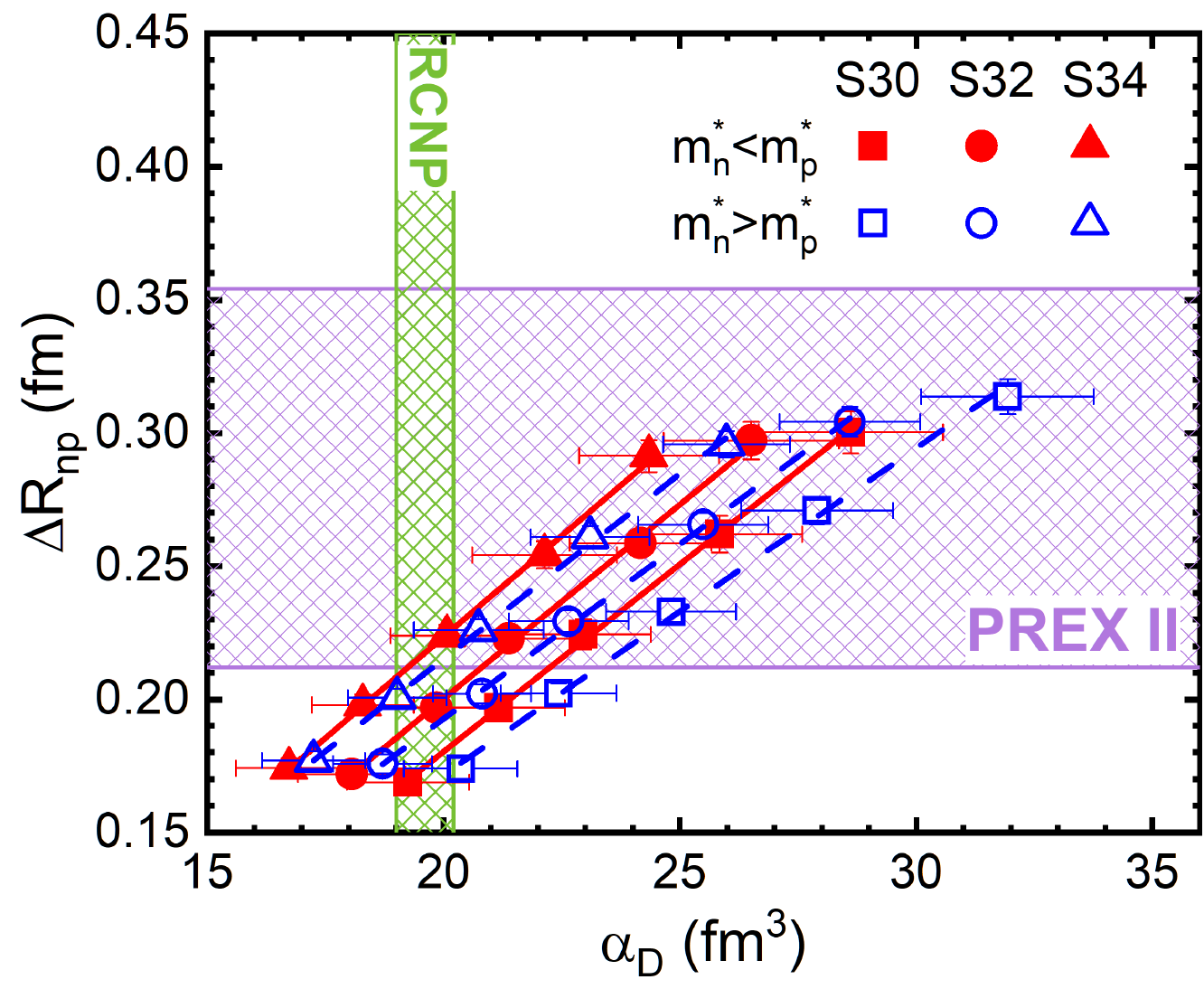}
	\caption{Correlation between the neutron-skin thickness \texorpdfstring{$\Delta R_{np}$}{Rnp} and the electric dipole polarizability \texorpdfstring{$\alpha_D$}{alphaD} for \texorpdfstring{$^{208}$Pb}{208Pb}. The error bars denote the corresponding standard deviation over 200 configurations.}
    \label{fig:Rnp-alphaD}
\end{figure}

To extract the favored effective interactions by simultaneously describing both $\Delta R_{np}$ and $\alpha_D$, we present $\Delta R_{np}$ as a function of $\alpha_D$ in Fig.~\ref{fig:Rnp-alphaD}. A clear linear correlation between these two observables is observed when $L$ is varied while keeping $S_0$ and the effective-mass splitting fixed, in agreement with the findings of Ref.~\cite{Reinhard10PRC,ZZLi21PRC}. 
Our calculations indicate that a simultaneous reproduction of $\Delta R_{np}$ and $\alpha_D$ favors relatively large values of the symmetry energy at saturation. If the initial-configuration dependence of $\alpha_D$ is considered as a part of theoretical uncertainties, the favored parameter region corresponds to $S_0 \simeq 32$-34~MeV and $L \simeq 66$-87~MeV for $f_I=0.19$ ($ m_n^*<m_p^*$), while for $f_I=-0.26$ ($m_n^*>m_p^*$) the favored region corresponds to $S_0 \simeq 34$~MeV with $L \simeq 64$-78~MeV.
The allowed $L$ intervals are determined by finding the overlap between the theoretical $\alpha_D$-$\Delta R_{np}$ band (including the error bar) and the overlap region between the Research Center for Nuclear Physics (RCNP) and PREX-II data bands.
It should be emphasized that $\alpha_D$ is predominantly sensitive to the symmetry energy at subsaturation densities, which means the extracted parameters $S_0$ and $L$ involve uncontrolled extrapolation. 


\begin{figure}[htbp]
	\centering
	\includegraphics[width=1\linewidth]{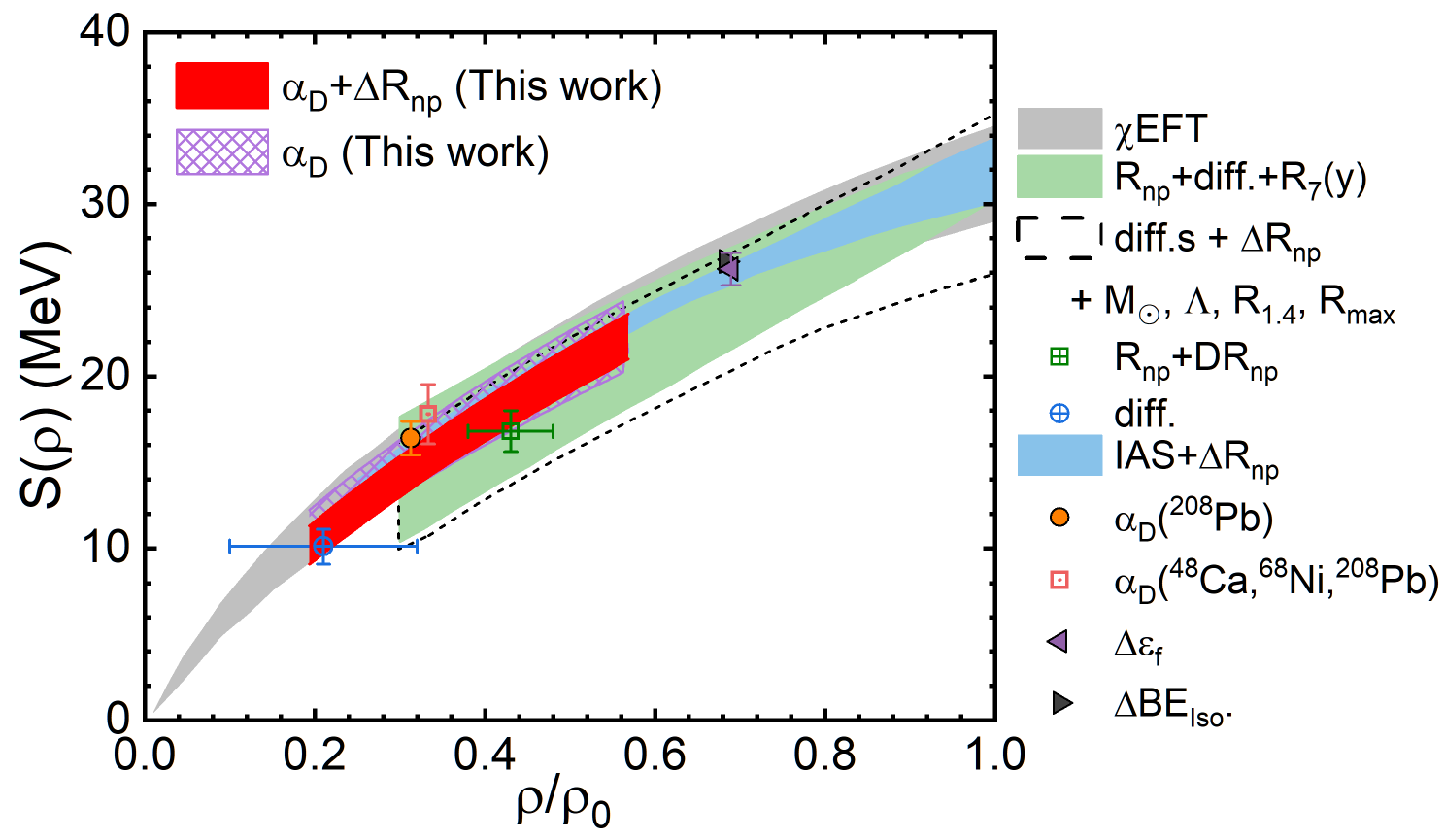}
	\caption{Constraints on the symmetry energy as a function of normalized density \texorpdfstring{$\rho/\rho_0$}{rho}.}
    \label{fig:rho-S0}
\end{figure}

It is therefore crucial to identify the density region to which the observables $\alpha_D$ and $\Delta R_{np}$ are most sensitive, and to extract the symmetry energy at the corresponding densities. Following the same procedure as in Ref.~\cite{ZZhang15PRC}, we evaluate the Pearson correlation coefficient between the inverse dipole polarizability $1/\alpha_D$ and the symmetry energy $S(\rho)$. We find that the correlation coefficient exceeds 0.99 over the subsaturation density interval $0.20 \le \rho/\rho_0 \le 0.57$. This strong correlation demonstrates that the IVGDR predominantly probes the symmetry energy in this specific density region. The constraint imposed solely by the RCNP $\alpha_D$ data is depicted by the purple region in Fig.~\ref{fig:rho-S0}. 
The constraint imposed simultaneously by the RCNP $\alpha_D$ and the PREX-II neutron-skin data is depicted by the red region in Fig.~\ref{fig:rho-S0}. 
Quantitatively, we obtain a tightly constrained symmetry energy of $S(\rho=0.2\rho_0)=10.18\pm1.10$ MeV, and $S(\rho=0.57\rho_0)=22.31\pm1.32$ MeV.

To provide a comprehensive comparison of symmetry-energy constraints at their respective sensitive densities, we present in Fig.~\ref{fig:rho-S0} a summary plot similar to that in Ref.~\cite{MYQiu25PRC}. Various isospin-sensitive observables that have been used to constrain the symmetry energy at subsaturation densities are included for comparison.
For instance, analyses of the electric dipole polarizability with RPA/QRPA (quasiparticle RPA) method provide constraints around $\rho/\rho_0\approx1/3$, yielding $S(\rho_0/3)=15.5$-17.4~MeV~\cite{JXu20PLB} and $16.9$-18.9~MeV~\cite{MYQiu25PRC}. The purple left-pointing triangle denotes the constraint extracted from the neutron-proton Fermi-energy difference at a subsaturation density of $\rho\simeq0.11~\mathrm{fm}^{-3}$, giving $S(\rho)=26.2\pm1.0$~MeV~\cite{NWang13PRC}. The dark-gray right-pointing triangle corresponds to the constraint deduced from binding energy differences between heavy isotope pairs, which yields $S(\rho_c)=26.65\pm0.20$~MeV at $\rho_c\approx0.11~\mathrm{fm}^{-3}$~\cite{ZZhang13PLB}. The isospin diffusion in HICs~\cite{Tsang09PRL,YXZhang20PRC,Lynch22PLB} constrains the symmetry energy at $\rho/\rho_0=0.21\pm0.11$ with $S(\rho)=10.1\pm1.0$~MeV. The single and double neutron-proton yield ratios constrain the symmetry energy near $\rho/\rho_0\approx0.43$ with $S(\rho)=16.8\pm1.2$~MeV~\cite{Morfouace19PLB}. 
Quantitatively, our extracted constraint on the symmetry energy at $\rho_0/3$ is 16.9$\pm$1.7 MeV, which is consistent with the RPA constraints~\cite{MYQiu25PRC,JXu20PLB} within the uncertainties. Within theoretical uncertainties, the constraints on the symmetry energy in the sensitive density region $0.20\le\rho/\rho_0\le0.57$ are consistent with those obtained via different approaches, which is narrower than the symmetry energy region predicted using chiral effective field theory~\cite{Drischler16PRC,Ciampi25PLB}. 

\subsection{Impacts of spin-orbit interaction on the constraints of symmetry energy}\label{SubSec:uncertainties}


Since the present AMD model does not include the spin orbit interaction, it is essential to estimate its impact on the extracted constraints on the symmetry energy. To this end, we select the Skyrme-Hartree-Fock-Bogolyubov approach~\cite{Bennaceur05CPC}, which is likewise based on fully antisymmetrized many-body wave functions and determines the ground state through variational energy minimization. Although the two frameworks differ in their treatment of correlations and clustering degrees of freedom, they are known to provide comparable bulk ground-state properties, particularly for density distributions, root-mean-square radii, and $\alpha_D$. This justifies the use of SHFB as a reference to estimate the missing spin-orbit effect in AMD. Within the SHFB framework, the spin-orbit interaction can be explicitly switched on and off, allowing for a controlled and quantitative assessment. We perform calculations using 30 Skyrme parametrization sets to reduce possible model dependence. The results consistently show that the impact of the spin-orbit interaction on the neutron-skin thickness $\Delta R_{np}$ is less than $2\%$. This finding is in line with previous studies indicating that $\Delta R_{np}$ is primarily governed by the isovector channel of the nuclear energy density functional and exhibits only a weak sensitivity to the spin-orbit strength, especially in heavy nuclei such as $^{208}\mathrm{Pb}$~\cite{LWChen10PRC,ZZhang13PLB}.

We further investigate the influence of the spin-orbit interaction on the electric dipole polarizability $\alpha_D$ within the SHFB+QRPA framework. The calculations show that turning off the spin-orbit interaction leads to a small increase of $\alpha_D$, and the difference between the two results is less than 3.9\%~\cite{ZZLi_private}. Since $\alpha_D$ is strongly correlated with $\Delta R_{np}$ and the symmetry energy slope parameter $L$, this result provides an additional consistency check on the impact of the spin-orbit interaction on the observables we analyzed.

Propagating these uncertainties to the extracted constraints on the symmetry energy parameters using the established correlations in Fig.~\ref{fig:Rnp-alphaD},  we find that the allowed range of $S_0$ is about $5\%$. For $L$, the changes in the lower and upper boundaries are about $3\%$ and $8\%$, respectively. 
Therefore, the absence of the spin-orbit interaction in AMD is not expected to qualitatively affect the present constraints.


\section{Summary}\label{Sec:summary}


In this work, we investigated the isovector giant dipole resonance and the neutron skin properties of $^{208}\mathrm{Pb}$ within the framework of the antisymmetrized molecular dynamics model. Our calculations show that the AMD model can naturally reproduce the strength function distribution $S(E)$ even without introducing additional smoothing parameters and two-body nucleon-nucleon collisions. 
This behavior can be attributed to an intrinsic Landau-damping mechanism in the transport models, in which the different local mean fields and effective masses experienced by individual Gaussian wave packets lead to different oscillation frequencies, and this phase-mixing mechanism leads to a gradual loss of phase coherence. 

Furthermore, our calculations show that the main frequency of the IVGDR is dominated by the symmetry potential at the surface of the nucleus. The main frequency of IVGDR increases with the symmetry energy coefficient $S_0$ and the strength of nucleon effective-mass splitting $\Delta m_{np}^*$, but decreases with $L$ because the symmetry energy at subsaturation density decreases with increasing $L$. 
As an $S(E)$-weighted integral observable, the electric dipole polarizability $\alpha_D$ depends on $S_0$, $L$, and $\Delta m_{np}^*$. In addition, the density profile and neutron skin of $^{208}\mathrm{Pb}$ can also be well described by AMD model with certain effective interaction parameters. The neutron-skin thickness $\Delta R_{np}$ is strongly sensitive to the slope of symmetry energy but weakly sensitive to the symmetry energy coefficient $S_0$ and nucleon effective-mass splitting $\Delta m_{np}^*$.

By performing a combined analysis of $\Delta R_{np}$ and $\alpha_D$ within the unified AMD framework, we have extracted favored symmetry energy parameters that are consistent with both the PREX-II neutron skin and RCNP dipole polarizability data. The combined constraints favor a relatively large symmetry energy at saturation, $S_0 \simeq 32$-34 MeV, and a slope parameter in the range $L \simeq 64$-87 MeV, with only a weak dependence on the effective-mass splitting. Furthermore, a correlation analysis reveals that the IVGDR predominantly probes the symmetry energy in the subsaturation density window $0.20 \le \rho/\rho_0 \le 0.57$. Within this region, we extract tight constraints on the density dependence of the symmetry energy, yielding $S(\rho=0.2\rho_0)=10.18\pm1.10$ MeV, and $S(\rho=0.57\rho_0)=22.31\pm1.32$ MeV.

In the future, working with other microscopic approaches such as energy density functional method will be essential to further reduce the uncertainties in the symmetry energy and to achieve a more consistent description across different theoretical frameworks and experimental observables.

\section*{Acknowledgments}
The authors thank valuable discussions with Professor L.W.Chen, Professor Y.F. Niu and Dr. Z.Z. Li. This work was supported by the National Natural
Science Foundation of China under Grants No. 12275359, No. 12375129, No. 11875323 and No. 11961141003, by JSPS KAKENHI Grant No. JP21K03528, by the National Key R\&D Program of China under Grant No. 2023 YFA1606402, by the Continuous Basic Scientific Research Project, by funding of the China Institute of Atomic Energy under Grant No. YZ222407001301, No. YZ232604001601, No. YC010270525794 and No. PA010271225779, and by the Leading Innovation Project of the CNNC under Grants No. LC192209000701 and No. LC202309000201. We acknowledge support by the computing server SCATP in China Institute of Atomic Energy.

\section*{DATA AVAILABILITY}
The data that support the findings of this article are not publicly available. The data are available from the authors
upon reasonable request.


\appendix
\section{Details of Parameter Sets}\label{Appendix:ParameterSets}

In Table~\ref{tab:parameters_x3_combined}, we list the detailed values of the parameter $x_3$ and $x_3'$ used in our calculations.
\begin{table}[htbp]
  \centering
  \footnotesize
  \caption{The parameter sets (\texorpdfstring{$S_0$}{S0}, \texorpdfstring{$L$}{L}) and corresponding coupling constants (\texorpdfstring{$x_3, x'_3$}{x3}) used in the calculations. The symbols \texorpdfstring{$\dagger$}{dagger} and \texorpdfstring{$\ddagger$}{ddagger} denote the SLy4 and SkM* interactions, respectively.}
  \label{tab:parameters_x3_combined}
  \setlength{\tabcolsep}{6pt}
  \renewcommand{\arraystretch}{1.15}
  \begin{tabular}{c c c c c c c}
    \hline
    \hline
   Sets & $m_s^*$ & $m_v^*$ & $S_0$ (MeV) & $L$ (MeV) & $x_3$ & $x'_3$ \\
    \hline
   1 & &    & &46  &       & $-$0.203 \\
   2 & &    & &61  &       & $-$0.653 \\
   3 & 0.695& 0.800  & 30 & 75  & 1.383 & $-$1.063 \\
   4 & &    & &92  &       & $-$1.573 \\
   5 & &    & &108 &       & $-$2.043 \\
    \hline
   6$\dagger$ & &    & &46  &       & 0.000 \\
   7 & &    & &61  &       & $-$0.444 \\
   8& 0.695& 0.800  & 32 & 75  & 1.354 & $-$0.854 \\
   9 & &    & &92  &       & $-$1.364 \\
   10 & &    & &108 &       & $-$1.854 \\
    \hline
   11 & &    & &46  &       & 0.206 \\
   12 & &    & &61  &       & $-$0.234 \\
   13 & 0.695& 0.800  & 34 & 75  & 1.324 & $-$0.664 \\
   14 & &    & &92  &       & $-$1.154 \\
   15 & &    & &108 &       & $-$1.644 \\
    \hline
    \hline
    16$\ddagger$ & &    & &46  &       & 0.000 \\
    17 & &    & &61  &       & $-0.390$ \\
    18 & 0.789 & 0.653 & 30 & 75  & 0.000 & $-0.770$ \\
    19 & &    & &92  &       & $-1.200$ \\
    20 & &    & &108 &       & $-1.630$ \\
    \hline
    21 & &    & &46  &       & 0.173 \\
    22 & &    & &61  &       & $-0.224$ \\
    23 &0.789 & 0.653 & 32 & 75  & $-0.026$ & $-0.584$ \\
    24 & &    & &92  &       & $-1.024$ \\
    25 & &    & &108 &       & $-1.444$ \\
    \hline
    26 & &    & &46  &       & 0.352 \\
    27 & &    & &61  &       & $-0.038$ \\
    28 &0.789 & 0.653  & 34 & 75  & $-0.052$ & $-0.398$ \\
    29 & &    & &92  &       & $-0.848$ \\
    30 & &    & &108 &       & $-1.258$ \\
    \hline
    \hline
  \end{tabular}
\end{table}
\section{Motion of dipole moment}\label{app:ddotD}

The dipole moment $\mathbf{D}$ is defined as the difference between the center-of-mass of protons and neutrons, i.e.,
\begin{equation}
    \begin{aligned}
        \mathbf{D}&\equiv e\frac{NZ}{A}(\mathbf{R}_{p}(t)-\mathbf{R}_{n}(t))\\
        &=e\frac{NZ}{A}(\frac{1}{Z}\sum_{i\in p}\mathbf{r}_{i}-\frac{1}{N}\sum_{i\in n}\mathbf{r}_{i}).
    \end{aligned}
\end{equation}
When a spherical nucleus is not perturbed by the dipole electric field, we will have $\mathbf{D}=0$. Under the dipole perturbation, $\mathbf{D}\ne 0$. If we move all neutrons (protons) by the same amount, i.e.,$\mathbf{r}_i\to \mathbf{r}_i-\mathbf{D}_n$ for neutrons and $\mathbf{r}_i\to \mathbf{r}_i-\mathbf{D}_p$ for protons, we have $\mathbf{D}=\mathbf{D}_p-\mathbf{D}_n$. 

Here we assume a Goldhaber-Teller-like mode, where proton and neutron densities undergo small rigid displacements. To linear order in the dipole coordinate, the local isospin asymmetry can be written as
\begin{equation}
    \begin{aligned}
        &\rho_n(\mathbf{r}_i)\to \rho_n(\mathbf{r}_i-\mathbf{D}_n)\approx \rho_n(\mathbf{r})-\nabla \rho_n\cdot \mathbf{D}_n,\\
        &\rho_p(\mathbf{r}_i)\to \rho_p(\mathbf{r}_i-\mathbf{D}_p)\approx \rho_p(\mathbf{r})-\nabla \rho_p\cdot \mathbf{D}_p.\\
    \end{aligned}
\end{equation}
Usually, the isospin asymmetry $\delta$ of the nucleus is relatively small and it means that we can approximately describe $\rho_n\approx\rho_p\approx\rho/2$. Thus, the total density remained almost constant because
\begin{equation}
    \begin{aligned}
        &\rho_n(\mathbf{r}_i)+\rho_p(\mathbf{r}_i)\to \rho_n(\mathbf{r}_i-\mathbf{D}_n)+\rho_p(\mathbf{r}_i-\mathbf{D}_p)\approx \rho(\mathbf{r}).\\
    \end{aligned} 
\end{equation}
In the above derivation, we also use the condition that $(\mathbf{D}_n+\mathbf{D}_p)=0$. Another, the difference between the neutron and proton densities becomes
\begin{equation}\label{eq:rhoD}
    \begin{aligned}
        &\rho_n(\mathbf{r}_i)-\rho_p(\mathbf{r}_i)\approx \frac{1}{2}\nabla \rho\cdot \mathbf{D}.\\
    \end{aligned} 
\end{equation}
It means that the isospin asymmetry can be expressed as
\begin{equation}
    \delta(\mathbf{r}_i)\approx\frac{1}{2}\frac{\nabla \rho}{\rho}\cdot \mathbf{D}
\end{equation}
and $\delta$ gives a contribution proportional to $\mathbf{D}$ in the expression for the mean field.

Now, let's see how $\mathbf{D}$ evolves. Based on the single-particle potential of the neutron and proton, the second-order time derivative of $\mathbf{D}$ becomes
\begin{equation}
    \begin{aligned}\label{eq:ddotD}
        \ddot{\mathbf{D}}&=e\frac{NZ}{A}\left(\frac{1}{Z}\sum_{i\in p}\ddot{x_i}-\frac{1}{N}\sum_{i\in n}\ddot{x_i}\right)\\
        &=e\frac{NZ}{A}\left(\frac{1}{Zm}\sum_{i\in p}m\ddot{x_i}-\frac{1}{Nm}\sum_{i\in n}m\ddot{x_i}\right)\\
        &=e\frac{NZ}{A}\left(\frac{1}{Zm}\sum_{i\in p}(-\nabla_i \epsilon_p )-\frac{1}{Nm}\sum_{i\in n}(-\nabla_i \epsilon_n )\right)\\
    \end{aligned}
\end{equation}
Here, $\epsilon_n$ and $\epsilon_p$ are the single-particle energy, which consists of the single-particle potential of the neutron and proton, i.e., $U_n$ and $U_p$, and the single-particle kinetic energy, $\epsilon_k^n$ and $\epsilon_k^p$.
\begin{equation}
    \begin{aligned}\label{eq:Usym}
        \epsilon_n=\frac{p^2}{2m_n^*}+U^{loc}_n\approx \frac{p_{F,n}^2}{2m_n^*}+U^{loc}_0+U^{loc}_{sym}\delta,\\
        \epsilon_p=\frac{p^2}{2m_p^*}+U^{loc}_p\approx \frac{p_{F,p}^2}{2m_p^*}+U^{loc}_0-U^{loc}_{sym}\delta.
    \end{aligned}
\end{equation}
$U^{loc}_0$ and $U_{sym}^{loc}$ mean that the potential arises from the local interactions for symmetric and asymmetric part, respectively. The nonlocal interactions are included in the effective kinetic-energy terms. 

By inserting the above formula into Eq.~(\ref{eq:ddotD}) and supposing that each nucleon feels the same single particle potential, we will have
\begin{align}
    \ddot{\mathbf{D}} & = -e\frac{NZ}{Am}\left(\nabla \epsilon_p -\nabla \epsilon_n\right) \nonumber\\
    &\approx e\frac{NZ}{Am} \nabla (\frac{p_{F,n}^2}{2m_n^*}-\frac{p_{F,p}^2}{2m_p^*}+2U^{loc}_{sym}\delta)\nonumber\\
    &= e\frac{NZ}{Am} \nabla \bigg( (\frac{\hbar^2}{2m_n^*}-\frac{\hbar^2}{2m_p^*})k_F^2\nonumber\\
    &\quad\quad +\frac{2}{3}(\frac{\hbar^2}{2m_n^*}+\frac{\hbar^2}{2m_p^*})k_F^2\delta +2U^{loc}_{sym}\delta\bigg)\nonumber\\
    &= e\frac{NZ}{Am} \nabla \left( (\frac{1}{m_s^*}-\frac{1}{m_v^*})\hbar^2k_F^2\delta+\frac{4}{3}\frac{\hbar^2}{2m_s^*}k_F^2\delta +2U^{loc}_{sym}\delta\right)\nonumber\\ 
    &= e\frac{NZ}{Am} \nabla \left( (\frac{1}{m_s^*}-\frac{1}{m_v^*})\hbar^2k_F^2\delta+4S(\rho)\delta\right)\nonumber\\   
\end{align}
In the above derivations, the following relationship is used. (i) The Laplacian term $\nabla^2\rho$ is approximately expressed in terms of $(\nabla\rho)^2/\rho$ in the nuclear surface region, where the density exhibits a Fermi-type profile. (ii) The $ \frac{1}{m_n^*}\frac{1}{m_p^*}\approx \frac{1}{m_s^{*2}}$ is used according to the relationship in Ref.~\cite{LWChen09PRC}. 
(iii) The density dependence of the symmetry energy can be expressed as~\cite{CXu10PRC}
\begin{equation}
    S(\rho)=\frac{1}{3}\frac{\hbar^2 k_F^2}{2m_N^*} +\frac{1}{2}U^{loc}_{sym}.
\end{equation}
Thus, we can rewrite the $\ddot{\mathbf{D}}$ as
\begin{equation}\label{eq:ddots-D}
    \begin{aligned}
           \ddot{\mathbf{D}}  &= e\frac{NZ}{Am} \nabla \left( (\frac{1}{m_s^*}-\frac{1}{m_v^*})\hbar^2k_F^2\delta+4S(\rho)\delta\right)\\\\
           &=e\frac{NZ}{Am} \nabla \left( (\frac{1}{m_s^*}-\frac{1}{m_v^*})\hbar^2k_F^2\frac{1}{2}\frac{\nabla \rho}{\rho}\cdot\mathbf{D}+4S(\rho)\frac{1}{2}\frac{\nabla \rho}{\rho}\cdot\mathbf{D} \right)\\    
           &\approx    e\frac{NZ}{Am} \nabla\bigg((\frac{1}{m_s^*}-\frac{1}{m_v^*})\hbar^2k_F^2\bigg)\frac{1}{2}\frac{\nabla\rho}{\rho}\cdot \mathbf{D}\\
           &\quad+e\frac{NZ}{Am} \bigg[  2\nabla S(\rho)\cdot\frac{\nabla\rho }{\rho} +2S(\rho)(-\frac{(\nabla\rho)^2}{\rho^2}+\frac{\nabla^2\rho}{\rho})\bigg] \cdot \mathbf{D}\\
           &=e\frac{NZ}{Am} \nabla\bigg((\frac{1}{m_s^*}-\frac{1}{m_v^*})\hbar^2k_F^2\bigg)\frac{1}{2}\frac{\nabla\rho}{\rho}\cdot \mathbf{D}\\
           &\quad+e\frac{NZ}{Am}\bigg[ 2\frac{\partial S(\rho )}{\partial \rho}\frac{(\nabla\rho)^2 }{\rho} +2S(\rho)(-2\frac{(\nabla\rho)^2}{\rho^2})\bigg] \cdot \mathbf{D}\\
           &=e\frac{NZ}{Am} \nabla\bigg((\frac{1}{m_s^*}-\frac{1}{m_v^*})\hbar^2k_F^2\bigg)\frac{1}{2}\frac{\nabla\rho}{\rho}\cdot \mathbf{D}\\
           &\quad +e\frac{NZ}{Am} \bigg[ \frac{2}{3}L(\rho) -4S(\rho)\bigg] \frac{(\nabla\rho)^2}{\rho^2} \mathbf{D}\\
    \end{aligned}
\end{equation}
It means the oscillation frequency will depend on the values of
isoscalar effective mass, isovector effective mass, the slope of symmetry energy and the strength of symmetry energy at subsaturation density as the gradient term vanishes in the interior of the nucleus.

To isolate $\mathbf{D}$ in the first term of Eq.(\ref{eq:ddots-D}) as a form like $k\mathbf{D}$, we rewrite it as
\begin{equation}
    \begin{aligned}
        &e\frac{NZ}{Am} \nabla\bigg((\frac{1}{m_s^*}-\frac{1}{m_v^*})\hbar^2k_F^2\bigg)\frac{1}{2}\frac{\nabla\rho}{\rho}\cdot \mathbf{D}\bigg)\\
        &= e\frac{NZ}{Am} \bigg[\nabla\bigg((\frac{1}{m_s^*}-\frac{1}{m_v^*})\hbar^2k_F^2\frac{1}{2}\frac{\nabla\rho}{\rho}\bigg)\bigg]^{T} \mathbf{D}.\\
    \end{aligned}
\end{equation}
It means that the frequency of the IVGDR is also influenced by the isoscalar and isovector effective mass. 

\bibliography{References}

@article{BALi08PR,
title = {Recent progress and new challenges in isospin physics with heavy-ion reactions},
journal = {Physics Reports},
volume = {464},
number = {4},
pages = {113-281},
year = {2008},
issn = {0370-1573},
doi = {https://doi.org/10.1016/j.physrep.2008.04.005},
url = {https://www.sciencedirect.com/science/article/pii/S0370157308001269},
author = {Bao-An Li and Lie-Wen Chen and Che Ming Ko}
}

@article{ROCA18PPNP,
title = {Nuclear equation of state from ground and collective excited state properties of nuclei},
journal = {Progress in Particle and Nuclear Physics},
volume = {101},
pages = {96-176},
year = {2018},
issn = {0146-6410},
doi = {https://doi.org/10.1016/j.ppnp.2018.04.001},
url = {https://www.sciencedirect.com/science/article/pii/S0146641018300334},
author = {X. Roca-Maza and N. Paar}
}

@article{Trippa08PRC,
  title = {Giant dipole resonance as a quantitative constraint on the symmetry energy},
  author = {Trippa, Luca and Col\`o, Gianluca and Vigezzi, Enrico},
  journal = {Phys. Rev. C},
  volume = {77},
  issue = {6},
  pages = {061304},
  numpages = {5},
  year = {2008},
  month = {Jun},
  publisher = {American Physical Society},
  doi = {10.1103/PhysRevC.77.061304},
  url = {https://link.aps.org/doi/10.1103/PhysRevC.77.061304}
}

@article{Roca13PRC,
  title = {Electric dipole polarizability in ${}^{208}$Pb: Insights from the droplet model},
  author = {Roca-Maza, X. and Brenna, M. and Col\`o, G. and Centelles, M. and Vi\~nas, X. and Agrawal, B. K. and Paar, N. and Vretenar, D. and Piekarewicz, J.},
  journal = {Phys. Rev. C},
  volume = {88},
  issue = {2},
  pages = {024316},
  numpages = {7},
  year = {2013},
  month = {Aug},
  publisher = {American Physical Society},
  doi = {10.1103/PhysRevC.88.024316},
  url = {https://link.aps.org/doi/10.1103/PhysRevC.88.024316}
}

@article{NWang13PRC,
  title = {Nuclear symmetry energy from the Fermi-energy difference in nuclei},
  author = {Wang, Ning and Ou, Li and Liu, Min},
  journal = {Phys. Rev. C},
  volume = {87},
  issue = {3},
  pages = {034327},
  numpages = {7},
  year = {2013},
  month = {Mar},
  publisher = {American Physical Society},
  doi = {10.1103/PhysRevC.87.034327},
  url = {https://link.aps.org/doi/10.1103/PhysRevC.87.034327}
}

@article{Tsang12PRC,
  title = {Constraints on the symmetry energy and neutron skins from experiments and theory},
  author = {Tsang, M. B. and Stone, J. R. and Camera, F. and Danielewicz, P. and Gandolfi, S. and Hebeler, K. and Horowitz, C. J. and Lee, Jenny and Lynch, W. G. and Kohley, Z. and Lemmon, R. and M\"oller, P. and Murakami, T. and Riordan, S. and Roca-Maza, X. and Sammarruca, F. and Steiner, A. W. and Vida\~na, I. and Yennello, S. J.},
  journal = {Phys. Rev. C},
  volume = {86},
  issue = {1},
  pages = {015803},
  numpages = {10},
  year = {2012},
  month = {Jul},
  publisher = {American Physical Society},
  doi = {10.1103/PhysRevC.86.015803},
  url = {https://link.aps.org/doi/10.1103/PhysRevC.86.015803}
}

@article{Reed21PRL,
  title = {Implications of PREX-2 on the Equation of State of Neutron-Rich Matter},
  author = {Reed, Brendan T. and Fattoyev, F. J. and Horowitz, C. J. and Piekarewicz, J.},
  journal = {Phys. Rev. Lett.},
  volume = {126},
  issue = {17},
  pages = {172503},
  numpages = {5},
  year = {2021},
  month = {Apr},
  publisher = {American Physical Society},
  doi = {10.1103/PhysRevLett.126.172503},
  url = {https://link.aps.org/doi/10.1103/PhysRevLett.126.172503}
}

@article{Reinhard10PRC,
  title = {Information content of a new observable: The case of the nuclear neutron skin},
  author = {Reinhard, P.-G. and Nazarewicz, W.},
  journal = {Phys. Rev. C},
  volume = {81},
  issue = {5},
  pages = {051303},
  numpages = {5},
  year = {2010},
  month = {May},
  publisher = {American Physical Society},
  doi = {10.1103/PhysRevC.81.051303},
  url = {https://link.aps.org/doi/10.1103/PhysRevC.81.051303}
}

@article{Piekarewicz14EPJ,
  title = {Symmetry energy constraints from giant resonances: A relativistic mean-field theory overview},
  volume = {50},
  ISSN = {1434-601X},
  url = {http://dx.doi.org/10.1140/epja/i2014-14025-x},
  DOI = {10.1140/epja/i2014-14025-x},
  number = {2},
  journal = {The European Physical Journal A},
  publisher = {Springer Science and Business Media LLC},
  author = {Piekarewicz,  J.},
  year = {2014},
  month = feb 
}

@article{CTao13PRC,
  title = {Pygmy and giant dipole resonances by Coulomb excitation using a quantum molecular dynamics model},
  volume = {87},
  ISSN = {1089-490X},
  url = {http://dx.doi.org/10.1103/PhysRevC.87.014621},
  DOI = {10.1103/physrevc.87.014621},
  number = {1},
  journal = {Physical Review C},
  publisher = {American Physical Society (APS)},
  author = {Tao,  C. and Ma,  Y. G. and Zhang,  G. Q. and Cao,  X. G. and Fang,  D. Q. and Wang,  H. W.},
  year = {2013},
  month = jan 
}

@article{TGYue22PRR,
  title = {Constraints on the symmetry energy from PREX-II in the multimessenger era},
  author = {Yue, Tong-Gang and Chen, Lie-Wen and Zhang, Zhen and Zhou, Ying},
  journal = {Phys. Rev. Res.},
  volume = {4},
  issue = {2},
  pages = {L022054},
  numpages = {6},
  year = {2022},
  month = {Jun},
  publisher = {American Physical Society},
  doi = {10.1103/PhysRevResearch.4.L022054},
  url = {https://link.aps.org/doi/10.1103/PhysRevResearch.4.L022054}
}

@article{Lynch22PLB,
title = {Decoding the density dependence of the nuclear symmetry energy},
journal = {Physics Letters B},
volume = {830},
pages = {137098},
year = {2022},
issn = {0370-2693},
doi = {https://doi.org/10.1016/j.physletb.2022.137098},
url = {https://www.sciencedirect.com/science/article/pii/S0370269322002325},
author = {W.G. Lynch and M.B. Tsang}
}

@article{Morfouace19PLB,
  title = {Constraining the symmetry energy with heavy-ion collisions and Bayesian analyses},
  volume = {799},
  ISSN = {0370-2693},
  url = {http://dx.doi.org/10.1016/j.physletb.2019.135045},
  DOI = {10.1016/j.physletb.2019.135045},
  journal = {Physics Letters B},
  publisher = {Elsevier BV},
  author = {Morfouace,  P. and Tsang,  C.Y. and Zhang,  Y. and Lynch,  W.G. and Tsang,  M.B. and Coupland,  D.D.S. and Youngs,  M. and Chajecki,  Z. and Famiano,  M.A. and Ghosh,  T.K. and Jhang,  G. and Lee,  Jenny and Liu,  H. and Sanetullaev,  A. and Showalter,  R. and Winkelbauer,  J.},
  year = {2019},
  month = dec,
  pages = {135045}
}

@article{JXu20PLB,
title = {Constraining isovector nuclear interactions with giant resonances within a Bayesian approach},
journal = {Physics Letters B},
volume = {810},
pages = {135820},
year = {2020},
issn = {0370-2693},
doi = {https://doi.org/10.1016/j.physletb.2020.135820},
url = {https://www.sciencedirect.com/science/article/pii/S0370269320306237},
author = {Jun Xu and Jia Zhou and Zhen Zhang and Wen-Jie Xie and Bao-An Li}
}

@article{Centelles09PRL,
  title = {Nuclear Symmetry Energy Probed by Neutron Skin Thickness of Nuclei},
  author = {Centelles, M. and Roca-Maza, X. and Vi\~nas, X. and Warda, M.},
  journal = {Phys. Rev. Lett.},
  volume = {102},
  issue = {12},
  pages = {122502},
  numpages = {4},
  year = {2009},
  month = {Mar},
  publisher = {American Physical Society},
  doi = {10.1103/PhysRevLett.102.122502},
  url = {https://link.aps.org/doi/10.1103/PhysRevLett.102.122502}
}

@article{MYQiu25PRC,
  title = {Symmetry energy and neutron matter equation of state at ${\ensuremath{\rho}}_{0}/3$ from the electric dipole polarizability in $^{48}\mathrm{Ca}$, $^{68}\mathrm{Ni}$, and $^{208}\mathrm{Pb}$},
  author = {Qiu, Mengying and Chen, Lie-Wen and Li, Zheng Zheng and Niu, Yi Fei and Zhang, Zhen},
  journal = {Phys. Rev. C},
  volume = {112},
  issue = {4},
  pages = {044312},
  numpages = {11},
  year = {2025},
  month = {Oct},
  publisher = {American Physical Society},
  doi = {10.1103/ssqr-62kc},
  url = {https://link.aps.org/doi/10.1103/ssqr-62kc}
}

@article{Tsang24nature,
  title = {Determination of the equation of state from nuclear experiments and neutron star observations},
  volume = {8},
  ISSN = {2397-3366},
  url = {http://dx.doi.org/10.1038/s41550-023-02161-z},
  DOI = {10.1038/s41550-023-02161-z},
  number = {3},
  journal = {Nature Astronomy},
  publisher = {Springer Science and Business Media LLC},
  author = {Tsang,  Chun Yuen and Tsang,  ManYee Betty and Lynch,  William G. and Kumar,  Rohit and Horowitz,  Charles J.},
  year = {2024},
  month = jan,
  pages = {328–336}
}

@article{Lattimer14EPJ,
  title = {Constraints on the symmetry energy using the mass-radius relation of neutron stars},
  volume = {50},
  ISSN = {1434-601X},
  url = {http://dx.doi.org/10.1140/epja/i2014-14040-y},
  DOI = {10.1140/epja/i2014-14040-y},
  number = {2},
  journal = {The European Physical Journal A},
  publisher = {Springer Science and Business Media LLC},
  author = {Lattimer,  James M. and Steiner,  Andrew W.},
  year = {2014},
  month = feb 
}

@article{Famiano06PRL,
  title = {Neutron and Proton Transverse Emission Ratio Measurements and the Density Dependence of the Asymmetry Term of the Nuclear Equation of State},
  author = {Famiano, M. A. and Liu, T. and Lynch, W. G. and Mocko, M. and Rogers, A. M. and Tsang, M. B. and Wallace, M. S. and Charity, R. J. and Komarov, S. and Sarantites, D. G. and Sobotka, L. G. and Verde, G.},
  journal = {Phys. Rev. Lett.},
  volume = {97},
  issue = {5},
  pages = {052701},
  numpages = {4},
  year = {2006},
  month = {Aug},
  publisher = {American Physical Society},
  doi = {10.1103/PhysRevLett.97.052701},
  url = {https://link.aps.org/doi/10.1103/PhysRevLett.97.052701}
}

@article{ZGXiao09PRL,
  title = {Circumstantial Evidence for a Soft Nuclear Symmetry Energy at Suprasaturation Densities},
  volume = {102},
  ISSN = {1079-7114},
  url = {http://dx.doi.org/10.1103/PhysRevLett.102.062502},
  DOI = {10.1103/physrevlett.102.062502},
  number = {6},
  journal = {Physical Review Letters},
  publisher = {American Physical Society (APS)},
  author = {Xiao,  Zhigang and Li,  Bao-An and Chen,  Lie-Wen and Yong,  Gao-Chan and Zhang,  Ming},
  year = {2009},
  month = feb 
}

@article{Cozma16PLB,
title = {The impact of energy conservation in transport models on the π−/π+ multiplicity ratio in heavy-ion collisions and the symmetry energy},
journal = {Physics Letters B},
volume = {753},
pages = {166-172},
year = {2016},
issn = {0370-2693},
doi = {https://doi.org/10.1016/j.physletb.2015.12.015},
url = {https://www.sciencedirect.com/science/article/pii/S0370269315009600},
author = {M.D. Cozma}
}

@article{Ciampi25PRC,
  title = {Model-independent measurement of isospin diffusion in Ni-Ni systems at intermediate energy},
  author = {Ciampi, C. and Frankland, J. D. and Gruyer, D. and Le Neindre, N. and Mallik, S. and Bougault, R. and Chbihi, A. and Baldesi, L. and Barlini, S. and Bonnet, E. and Borderie, B. and Camaiani, A. and Casini, G. and Dekhissi, I. and Dell'Aquila, D. and Due\~nas, J. A. and Fable, Q. and Gramegna, F. and Gouyet, C. and Henri, M. and Hong, B. and Kim, S. and Kordyasz, A. and Kozik, T. and Kweon, M. J. and Lombardo, I. and Lopez, O. and Manduci, L. and Marchi, T. and Mazurek, K. and Nam, S. H. and Park, J. and P\^arlog, M. and Pasquali, G. and Piantelli, S. and Poggi, G. and Rebillard-Souli\'e, A. and Revenko, R. and Valdr\'e, S. and Verde, G. and Vient, E.},
  collaboration = {INDRA-FAZIA Collaboration},
  journal = {Phys. Rev. C},
  volume = {111},
  issue = {4},
  pages = {044601},
  numpages = {12},
  year = {2025},
  month = {Apr},
  publisher = {American Physical Society},
  doi = {10.1103/PhysRevC.111.044601},
  url = {https://link.aps.org/doi/10.1103/PhysRevC.111.044601}
}

@article{NBZhang18APJ,
doi = {10.3847/1538-4357/aac027},
url = {https://doi.org/10.3847/1538-4357/aac027},
year = {2018},
month = {may},
publisher = {The American Astronomical Society},
volume = {859},
number = {2},
pages = {90},
author = {Zhang, Nai-Bo and Li, Bao-An and Xu, Jun},
title = {Combined Constraints on the Equation of State of Dense Neutron-rich Matter from Terrestrial Nuclear Experiments and Observations of Neutron Stars},
journal = {The Astrophysical Journal}
}

@article{Tsang09PRL,
  title = {Constraints on the Density Dependence of the Symmetry Energy},
  author = {Tsang, M. B. and Zhang, Yingxun and Danielewicz, P. and Famiano, M. and Li, Zhuxia and Lynch, W. G. and Steiner, A. W.},
  journal = {Phys. Rev. Lett.},
  volume = {102},
  issue = {12},
  pages = {122701},
  numpages = {4},
  year = {2009},
  month = {Mar},
  publisher = {American Physical Society},
  doi = {10.1103/PhysRevLett.102.122701},
  url = {https://link.aps.org/doi/10.1103/PhysRevLett.102.122701}
}

@article{LWChen05PRL,
  title = {Determination of the Stiffness of the Nuclear Symmetry Energy from Isospin Diffusion},
  author = {Chen, Lie-Wen and Ko, Che Ming and Li, Bao-An},
  journal = {Phys. Rev. Lett.},
  volume = {94},
  issue = {3},
  pages = {032701},
  numpages = {4},
  year = {2005},
  month = {Jan},
  publisher = {American Physical Society},
  doi = {10.1103/PhysRevLett.94.032701},
  url = {https://link.aps.org/doi/10.1103/PhysRevLett.94.032701}
}

@article{ZYSun10PRC,
  title = {Isospin diffusion and equilibration for $\mathrm{Sn}+\mathrm{Sn}$ collisions at $E/A=35$ MeV},
  author = {Sun, Z. Y. and Tsang, M. B. and Lynch, W. G. and Verde, G. and Amorini, F. and Andronenko, L. and Andronenko, M. and Cardella, G. and Chatterje, M. and Danielewicz, P. and De Filippo, E. and Dinh, P. and Galichet, E. and Geraci, E. and Hua, H. and La Guidara, E. and Lanzalone, G. and Liu, H. and Lu, F. and Lukyanov, S. and Maiolino, C. and Pagano, A. and Piantelli, S. and Papa, M. and Pirrone, S. and Politi, G. and Porto, F. and Rizzo, F. and Russotto, P. and Santonocito, D. and Zhang, Y. X.},
  journal = {Phys. Rev. C},
  volume = {82},
  issue = {5},
  pages = {051603},
  numpages = {4},
  year = {2010},
  month = {Nov},
  publisher = {American Physical Society},
  doi = {10.1103/PhysRevC.82.051603},
  url = {https://link.aps.org/doi/10.1103/PhysRevC.82.051603}
}

@article{JXU20PRC,
  title = {Nucleus giant resonances from an improved isospin-dependent Boltzmann-Uehling-Uhlenbeck transport approach},
  author = {Xu, Jun and Qin, Wen-Tao},
  journal = {Phys. Rev. C},
  volume = {102},
  issue = {2},
  pages = {024306},
  numpages = {9},
  year = {2020},
  month = {Aug},
  publisher = {American Physical Society},
  doi = {10.1103/PhysRevC.102.024306},
  url = {https://link.aps.org/doi/10.1103/PhysRevC.102.024306}
}

@article{YYLiu21PRC,
  title = {Insights into the pion production mechanism and the symmetry energy at high density},
  author = {Liu, Yangyang and Wang, Yongjia and Cui, Ying and Xia, Cheng-Jun and Li, Zhuxia and Chen, Yongjing and Li, Qingfeng and Zhang, Yingxun},
  journal = {Phys. Rev. C},
  volume = {103},
  issue = {1},
  pages = {014616},
  numpages = {11},
  year = {2021},
  month = {Jan},
  publisher = {American Physical Society},
  doi = {10.1103/PhysRevC.103.014616},
  url = {https://link.aps.org/doi/10.1103/PhysRevC.103.014616}
}

@article{WBHe16PRC,
  title = {Dipole oscillation modes in light $\ensuremath{\alpha}$-clustering nuclei},
  author = {He, W. B. and Ma, Y. G. and Cao, X. G. and Cai, X. Z. and Zhang, G. Q.},
  journal = {Phys. Rev. C},
  volume = {94},
  issue = {1},
  pages = {014301},
  numpages = {12},
  year = {2016},
  month = {Jul},
  publisher = {American Physical Society},
  doi = {10.1103/PhysRevC.94.014301},
  url = {https://link.aps.org/doi/10.1103/PhysRevC.94.014301}
}

@article{HYKong17PRC,
  title = {Constraining simultaneously nuclear symmetry energy and neutron-proton effective mass splitting with nucleus giant resonances using a dynamical approach},
  author = {Kong, Hai-Yun and Xu, Jun and Chen, Lie-Wen and Li, Bao-An and Ma, Yu-Gang},
  journal = {Phys. Rev. C},
  volume = {95},
  issue = {3},
  pages = {034324},
  numpages = {9},
  year = {2017},
  month = {Mar},
  publisher = {American Physical Society},
  doi = {10.1103/PhysRevC.95.034324},
  url = {https://link.aps.org/doi/10.1103/PhysRevC.95.034324}
}

@article{Roca11PRL,
  title = {Neutron Skin of $^{208}\mathrm{Pb}$, Nuclear Symmetry Energy, and the Parity Radius Experiment},
  author = {Roca-Maza, X. and Centelles, M. and Vi\~nas, X. and Warda, M.},
  journal = {Phys. Rev. Lett.},
  volume = {106},
  issue = {25},
  pages = {252501},
  numpages = {4},
  year = {2011},
  month = {Jun},
  publisher = {American Physical Society},
  doi = {10.1103/PhysRevLett.106.252501},
  url = {https://link.aps.org/doi/10.1103/PhysRevLett.106.252501}
}

@article{ZZhang13PLB,
  title = {Constraining the symmetry energy at subsaturation densities using isotope binding energy difference and neutron skin thickness},
  volume = {726},
  ISSN = {0370-2693},
  url = {http://dx.doi.org/10.1016/j.physletb.2013.08.002},
  DOI = {10.1016/j.physletb.2013.08.002},
  number = {1–3},
  journal = {Physics Letters B},
  publisher = {Elsevier BV},
  author = {Zhang,  Zhen and Chen,  Lie-Wen},
  year = {2013},
  month = oct,
  pages = {234–238}
}

@article{YXZhang18PRC,
  title = {Comparison of heavy-ion transport simulations: Collision integral in a box},
  author = {Zhang, Ying-Xun and Wang, Yong-Jia and Colonna, Maria and Danielewicz, Pawel and Ono, Akira and Tsang, Manyee Betty and Wolter, Hermann and Xu, Jun and Chen, Lie-Wen and Cozma, Dan and Feng, Zhao-Qing and Das Gupta, Subal and Ikeno, Natsumi and Ko, Che-Ming and Li, Bao-An and Li, Qing-Feng and Li, Zhu-Xia and Mallik, Swagata and Nara, Yasushi and Ogawa, Tatsuhiko and Ohnishi, Akira and Oliinychenko, Dmytro and Papa, Massimo and Petersen, Hannah and Su, Jun and Song, Taesoo and Weil, Janus and Wang, Ning and Zhang, Feng-Shou and Zhang, Zhen},
  journal = {Phys. Rev. C},
  volume = {97},
  issue = {3},
  pages = {034625},
  numpages = {20},
  year = {2018},
  month = {Mar},
  publisher = {American Physical Society},
  doi = {10.1103/PhysRevC.97.034625},
  url = {https://link.aps.org/doi/10.1103/PhysRevC.97.034625}
}

@article{Roca13PRC-GQR,
  title = {Giant quadrupole resonances in ${}^{208}$Pb, the nuclear symmetry energy, and the neutron skin thickness},
  author = {Roca-Maza, X. and Brenna, M. and Agrawal, B. K. and Bortignon, P. F. and Col\`o, G. and Cao, Li-Gang and Paar, N. and Vretenar, D.},
  journal = {Phys. Rev. C},
  volume = {87},
  issue = {3},
  pages = {034301},
  numpages = {9},
  year = {2013},
  month = {Mar},
  publisher = {American Physical Society},
  doi = {10.1103/PhysRevC.87.034301},
  url = {https://link.aps.org/doi/10.1103/PhysRevC.87.034301}
}

@article{RWang20PLB,
title = {Constraining the in-medium nucleon-nucleon cross section from the width of nuclear giant dipole resonance},
journal = {Physics Letters B},
volume = {807},
pages = {135532},
year = {2020},
issn = {0370-2693},
doi = {https://doi.org/10.1016/j.physletb.2020.135532},
url = {https://www.sciencedirect.com/science/article/pii/S0370269320303361},
author = {Rui Wang and Zhen Zhang and Lie-Wen Chen and Che Ming Ko and Yu-Gang Ma}
}

@article{WBHe14PRL,
  title = {Giant Dipole Resonance as a Fingerprint of $\ensuremath{\alpha}$ Clustering Configurations in $^{12}\mathrm{C}$ and $^{16}\mathrm{O}$},
  author = {He, W. B. and Ma, Y. G. and Cao, X. G. and Cai, X. Z. and Zhang, G. Q.},
  journal = {Phys. Rev. Lett.},
  volume = {113},
  issue = {3},
  pages = {032506},
  numpages = {6},
  year = {2014},
  month = {Jul},
  publisher = {American Physical Society},
  doi = {10.1103/PhysRevLett.113.032506},
  url = {https://link.aps.org/doi/10.1103/PhysRevLett.113.032506}
}

@article{Reinhard21PRL,
  title = {Information Content of the Parity-Violating Asymmetry in $^{208}\mathrm{Pb}$},
  author = {Reinhard, Paul-Gerhard and Roca-Maza, Xavier and Nazarewicz, Witold},
  journal = {Phys. Rev. Lett.},
  volume = {127},
  issue = {23},
  pages = {232501},
  numpages = {7},
  year = {2021},
  month = {Nov},
  publisher = {American Physical Society},
  doi = {10.1103/PhysRevLett.127.232501},
  url = {https://link.aps.org/doi/10.1103/PhysRevLett.127.232501}
}

@article{Natsumi16PRC,
  title = {Probing neutron-proton dynamics by pions},
  author = {Ikeno, Natsumi and Ono, Akira and Nara, Yasushi and Ohnishi, Akira},
  journal = {Phys. Rev. C},
  volume = {93},
  issue = {4},
  pages = {044612},
  numpages = {13},
  year = {2016},
  month = {Apr},
  publisher = {American Physical Society},
  doi = {10.1103/PhysRevC.93.044612},
  url = {https://link.aps.org/doi/10.1103/PhysRevC.93.044612}
}

@article{Adhikari21PRL,
  title = {Accurate Determination of the Neutron Skin Thickness of $^{208}\mathrm{Pb}$ through Parity-Violation in Electron Scattering},
  author = {Adhikari, D. and Albataineh, H. and Androic, D. and Aniol, K. and Armstrong, D. S. and Averett, T. and Ayerbe Gayoso, C. and Barcus, S. and Bellini, V. and Beminiwattha, R. S. and Benesch, J. F. and Bhatt, H. and Bhatta Pathak, D. and Bhetuwal, D. and Blaikie, B. and Campagna, Q. and Camsonne, A. and Cates, G. D. and Chen, Y. and Clarke, C. and Cornejo, J. C. and Covrig Dusa, S. and Datta, P. and Deshpande, A. and Dutta, D. and Feldman, C. and Fuchey, E. and Gal, C. and Gaskell, D. and Gautam, T. and Gericke, M. and Ghosh, C. and Halilovic, I. and Hansen, J.-O. and Hauenstein, F. and Henry, W. and Horowitz, C. J. and Jantzi, C. and Jian, S. and Johnston, S. and Jones, D. C. and Karki, B. and Katugampola, S. and Keppel, C. and King, P. M. and King, D. E. and Knauss, M. and Kumar, K. S. and Kutz, T. and Lashley-Colthirst, N. and Leverick, G. and Liu, H. and Liyange, N. and Malace, S. and Mammei, R. and Mammei, J. and McCaughan, M. and McNulty, D. and Meekins, D. and Metts, C. and Michaels, R. and Mondal, M. M. and Napolitano, J. and Narayan, A. and Nikolaev, D. and Rashad, M. N. H. and Owen, V. and Palatchi, C. and Pan, J. and Pandey, B. and Park, S. and Paschke, K. D. and Petrusky, M. and Pitt, M. L. and Premathilake, S. and Puckett, A. J. R. and Quinn, B. and Radloff, R. and Rahman, S. and Rathnayake, A. and Reed, B. T. and Reimer, P. E. and Richards, R. and Riordan, S. and Roblin, Y. and Seeds, S. and Shahinyan, A. and Souder, P. and Tang, L. and Thiel, M. and Tian, Y. and Urciuoli, G. M. and Wertz, E. W. and Wojtsekhowski, B. and Yale, B. and Ye, T. and Yoon, A. and Zec, A. and Zhang, W. and Zhang, J. and Zheng, X.},
  collaboration = {PREX Collaboration},
  journal = {Phys. Rev. Lett.},
  volume = {126},
  issue = {17},
  pages = {172502},
  numpages = {7},
  year = {2021},
  month = {Apr},
  publisher = {American Physical Society},
  doi = {10.1103/PhysRevLett.126.172502},
  url = {https://link.aps.org/doi/10.1103/PhysRevLett.126.172502}
}

@article{YDSong23PRC,
  title = {In-medium nucleon-nucleon cross sections from characteristics of nuclear giant resonances and nuclear stopping power},
  author = {Song, Yi-Dan and Wang, Rui and Zhang, Zhen and Ma, Yu-Gang},
  journal = {Phys. Rev. C},
  volume = {108},
  issue = {6},
  pages = {064603},
  numpages = {10},
  year = {2023},
  month = {Dec},
  publisher = {American Physical Society},
  doi = {10.1103/PhysRevC.108.064603},
  url = {https://link.aps.org/doi/10.1103/PhysRevC.108.064603}
}

@article{Kanada05PRC,
  title = {Dipole resonances in light neutron-rich nuclei studied with time-dependent calculations of antisymmetrized molecular dynamics},
  author = {Kanada-En'yo, Y. and Kimura, M.},
  journal = {Phys. Rev. C},
  volume = {72},
  issue = {6},
  pages = {064301},
  numpages = {13},
  year = {2005},
  month = {Dec},
  publisher = {American Physical Society},
  doi = {10.1103/PhysRevC.72.064301},
  url = {https://link.aps.org/doi/10.1103/PhysRevC.72.064301}
}

@article{YXZhang14PLB,
title = {Constraints on nucleon effective mass splitting with heavy ion collisions},
journal = {Physics Letters B},
volume = {732},
pages = {186-190},
year = {2014},
issn = {0370-2693},
doi = {https://doi.org/10.1016/j.physletb.2014.03.030},
url = {https://www.sciencedirect.com/science/article/pii/S0370269314001865},
author = {Yingxun Zhang and M.B. Tsang and Zhuxia Li and Hang Liu}
}

@article{JPYang21PRC,
  title = {Influence of the treatment of initialization and mean-field potential on the neutron to proton yield ratios},
  author = {Yang, Junping and Zhang, Yingxun and Wang, Ning and Li, Zhuxia},
  journal = {Phys. Rev. C},
  volume = {104},
  issue = {2},
  pages = {024605},
  numpages = {9},
  year = {2021},
  month = {Aug},
  publisher = {American Physical Society},
  doi = {10.1103/PhysRevC.104.024605},
  url = {https://link.aps.org/doi/10.1103/PhysRevC.104.024605}
}

@article{ZZhang16PRC,
  title = {Isospin splitting of the nucleon effective mass from giant resonances in $^{208}\mathrm{Pb}$},
  author = {Zhang, Zhen and Chen, Lie-Wen},
  journal = {Phys. Rev. C},
  volume = {93},
  issue = {3},
  pages = {034335},
  numpages = {7},
  year = {2016},
  month = {Mar},
  publisher = {American Physical Society},
  doi = {10.1103/PhysRevC.93.034335},
  url = {https://link.aps.org/doi/10.1103/PhysRevC.93.034335}
}

@article{LWChen10PRC,
  title = {Density slope of the nuclear symmetry energy from the neutron skin thickness of heavy nuclei},
  author = {Chen, Lie-Wen and Ko, Che Ming and Li, Bao-An and Xu, Jun},
  journal = {Phys. Rev. C},
  volume = {82},
  issue = {2},
  pages = {024321},
  numpages = {7},
  year = {2010},
  month = {Aug},
  publisher = {American Physical Society},
  doi = {10.1103/PhysRevC.82.024321},
  url = {https://link.aps.org/doi/10.1103/PhysRevC.82.024321}
}

@article{YXZhang20PRC,
  title = {Constraints on the symmetry energy and its associated parameters from nuclei to neutron stars},
  author = {Zhang, Yingxun and Liu, Min and Xia, Cheng-Jun and Li, Zhuxia and Biswal, S. K.},
  journal = {Phys. Rev. C},
  volume = {101},
  issue = {3},
  pages = {034303},
  numpages = {11},
  year = {2020},
  month = {Mar},
  publisher = {American Physical Society},
  doi = {10.1103/PhysRevC.101.034303},
  url = {https://link.aps.org/doi/10.1103/PhysRevC.101.034303}
}

@article{Burgio21PPNP,
title = {Neutron stars and the nuclear equation of state},
journal = {Progress in Particle and Nuclear Physics},
volume = {120},
pages = {103879},
year = {2021},
issn = {0146-6410},
doi = {https://doi.org/10.1016/j.ppnp.2021.103879},
url = {https://www.sciencedirect.com/science/article/pii/S0146641021000338},
author = {G.F. Burgio and H.-J. Schulze and I. Vidaña and J.-B. Wei}
}

@misc{NuPECC2024,
  author       = {{NuPECC}},
  title        = {NuPECC Long Range Plan 2024 for European Nuclear Physics},
  year         = {2024},
  institution  = {Nuclear Physics European Collaboration Committee},
  url          = {https://www.nupecc.org/lrp2024/Documents/nupecc_lrp2024.pdf}
}

@article{LWChen09PRC,
  title = {Higher-order effects on the incompressibility of isospin asymmetric nuclear matter},
  author = {Chen, Lie-Wen and Cai, Bao-Jun and Ko, Che Ming and Li, Bao-An and Shen, Chun and Xu, Jun},
  journal = {Phys. Rev. C},
  volume = {80},
  issue = {1},
  pages = {014322},
  numpages = {24},
  year = {2009},
  month = {Jul},
  publisher = {American Physical Society},
  doi = {10.1103/PhysRevC.80.014322},
  url = {https://link.aps.org/doi/10.1103/PhysRevC.80.014322}
}

@article{Ciampi25PLB,
title = {Constraining the nuclear symmetry energy with Fermi-energy heavy ion collisions},
journal = {Physics Letters B},
volume = {868},
pages = {139815},
year = {2025},
issn = {0370-2693},
doi = {https://doi.org/10.1016/j.physletb.2025.139815},
url = {https://www.sciencedirect.com/science/article/pii/S0370269325005763},
author = {C. Ciampi and S. Mallik and F. Gulminelli and D. Gruyer and J.D. Frankland and N. {Le Neindre} and R. Bougault and A. Chbihi and L. Baldesi and S. Barlini and B. Borderie and A. Camaiani and G. Casini and I. Dekhissi and J.A. Dueñas and Q. Fable and F. Gramegna and M. Henri and B. Hong and S. Kim and A. Kordyasz and T. Kozik and I. Lombardo and O. Lopez and T. Marchi and S.H. Nam and J. Park and M. Pârlog and G. Pasquali and S. Piantelli and G. Poggi and S. Valdré and G. Verde and E. Vient}
}

@techreport{USDOE23LRP,
  author       = {US Department of Energy (USDOE)},
  title        = {A New Era of Discovery: The 2023 Long Range Plan for Nuclear Science},
  institution  = {US Department of Energy (USDOE), Washington, DC (United States). Office of Science},
  doi          = {10.2172/2280968},
  url          = {https://www.osti.gov/biblio/2280968},
  place        = {United States},
  year         = {2023},
  month        = {10}}

@article{BALi21Universe,
  title = {Progress in Constraining Nuclear Symmetry Energy Using Neutron Star Observables Since GW170817},
  volume = {7},
  ISSN = {2218-1997},
  url = {http://dx.doi.org/10.3390/universe7060182},
  DOI = {10.3390/universe7060182},
  number = {6},
  journal = {Universe},
  publisher = {MDPI AG},
  author = {Li,  Bao-An and Cai,  Bao-Jun and Xie,  Wen-Jie and Zhang,  Nai-Bo},
  year = {2021},
  month = jun,
  pages = {182}
}

@article{Ono92PRL,
  title = {Fragment formation studied with antisymmetrized version of molecular dynamics with two-nucleon collisions},
  author = {Ono, A. and Horiuchi, H. and Maruyama, T. and Ohnishi, A.},
  journal = {Phys. Rev. Lett.},
  volume = {68},
  issue = {19},
  pages = {2898--2900},
  numpages = {0},
  year = {1992},
  month = {May},
  publisher = {American Physical Society},
  doi = {10.1103/PhysRevLett.68.2898},
  url = {https://link.aps.org/doi/10.1103/PhysRevLett.68.2898}
}

@article{Ono92PTP,
    author = {Ono, Akira and Horiuchi, Hisashi and Maruyama, Toshiki and Ohnishi, Akira},
    title = {Antisymmetrized Version of Molecular Dynamics with Two-Nucleon Collisions and Its Application to Heavy Ion Reactions},
    journal = {Progress of Theoretical Physics},
    volume = {87},
    number = {5},
    pages = {1185-1206},
    year = {1992},
    month = {05},
    issn = {0033-068X},
    doi = {10.1143/ptp/87.5.1185},
    url = {https://doi.org/10.1143/ptp/87.5.1185},
    eprint = {https://academic.oup.com/ptp/article-pdf/87/5/1185/5272175/87-5-1185.pdf},
}

@article{Roca15PRC,
  title={Neutron skin thickness from the measured electric dipole polarizability in Ni 68, Sn 120, and Pb 208},
  author={Roca-Maza, Xavier and Vinas, Xavier and Centelles, Mario and Agrawal, BK and Colo, Gianluca and Paar, Nils and Piekarewicz, Jorge and Vretenar, Dario},
  journal={Physical Review C},
  volume={92},
  number={6},
  pages={064304},
  year={2015},
  publisher={APS}
}

@article{Drischler16PRC,
  title = {Asymmetric nuclear matter based on chiral two- and three-nucleon interactions},
  author = {Drischler, C. and Hebeler, K. and Schwenk, A.},
  journal = {Phys. Rev. C},
  volume = {93},
  issue = {5},
  pages = {054314},
  numpages = {12},
  year = {2016},
  month = {May},
  publisher = {American Physical Society},
  doi = {10.1103/PhysRevC.93.054314},
  url = {https://link.aps.org/doi/10.1103/PhysRevC.93.054314}
}

@article{Vretenar12PRC,
  title = {Low-energy isovector and isoscalar dipole response in neutron-rich nuclei},
  author = {Vretenar, D. and Niu, Y. F. and Paar, N. and Meng, J.},
  journal = {Phys. Rev. C},
  volume = {85},
  issue = {4},
  pages = {044317},
  numpages = {8},
  year = {2012},
  month = {Apr},
  publisher = {American Physical Society},
  doi = {10.1103/PhysRevC.85.044317},
  url = {https://link.aps.org/doi/10.1103/PhysRevC.85.044317}
}

@article{Tamii11PRL,
  title = {Complete Electric Dipole Response and the Neutron Skin in $^{208}\mathrm{Pb}$},
  author = {Tamii, A. and Poltoratska, I. and von Neumann-Cosel, P. and Fujita, Y. and Adachi, T. and Bertulani, C. A. and Carter, J. and Dozono, M. and Fujita, H. and Fujita, K. and Hatanaka, K. and Ishikawa, D. and Itoh, M. and Kawabata, T. and Kalmykov, Y. and Krumbholz, A. M. and Litvinova, E. and Matsubara, H. and Nakanishi, K. and Neveling, R. and Okamura, H. and Ong, H. J. and \"Ozel-Tashenov, B. and Ponomarev, V. Yu. and Richter, A. and Rubio, B. and Sakaguchi, H. and Sakemi, Y. and Sasamoto, Y. and Shimbara, Y. and Shimizu, Y. and Smit, F. D. and Suzuki, T. and Tameshige, Y. and Wambach, J. and Yamada, R. and Yosoi, M. and Zenihiro, J.},
  journal = {Phys. Rev. Lett.},
  volume = {107},
  issue = {6},
  pages = {062502},
  numpages = {5},
  year = {2011},
  month = {Aug},
  publisher = {American Physical Society},
  doi = {10.1103/PhysRevLett.107.062502},
  url = {https://link.aps.org/doi/10.1103/PhysRevLett.107.062502}
}

@article{Klimkiewicz07PRC,
  title = {Nuclear symmetry energy and neutron skins derived from pygmy dipole resonances},
  author = {Klimkiewicz, A. and Paar, N. and Adrich, P. and Fallot, M. and Boretzky, K. and Aumann, T. and Cortina-Gil, D. and Pramanik, U. Datta and Elze, Th. W. and Emling, H. and Geissel, H. and Hellstr\"om, M. and Jones, K. L. and Kratz, J. V. and Kulessa, R. and Nociforo, C. and Palit, R. and Simon, H. and Sur\'owka, G. and S\"ummerer, K. and Vretenar, D. and Walu\ifmmode \acute{s}\else \'{s}\fi{}, W.},
  collaboration = {LAND Collaboration},
  journal = {Phys. Rev. C},
  volume = {76},
  issue = {5},
  pages = {051603},
  numpages = {4},
  year = {2007},
  month = {Nov},
  publisher = {American Physical Society},
  doi = {10.1103/PhysRevC.76.051603},
  url = {https://link.aps.org/doi/10.1103/PhysRevC.76.051603}
}

@article{Carbone10PRC,
  title = {Constraints on the symmetry energy and neutron skins from pygmy resonances in $^{68}\mathrm{Ni}$ and $^{132}\mathrm{Sn}$},
  author = {Carbone, Andrea and Col\`o, Gianluca and Bracco, Angela and Cao, Li-Gang and Bortignon, Pier Francesco and Camera, Franco and Wieland, Oliver},
  journal = {Phys. Rev. C},
  volume = {81},
  issue = {4},
  pages = {041301},
  numpages = {5},
  year = {2010},
  month = {Apr},
  publisher = {American Physical Society},
  doi = {10.1103/PhysRevC.81.041301},
  url = {https://link.aps.org/doi/10.1103/PhysRevC.81.041301}
}

@article{Colo14EPJ,
  title = {Symmetry energy from the nuclear collective motion: constraints from dipole,  quadrupole,  monopole and spin-dipole resonances},
  volume = {50},
  ISSN = {1434-601X},
  url = {http://dx.doi.org/10.1140/epja/i2014-14026-9},
  DOI = {10.1140/epja/i2014-14026-9},
  number = {2},
  journal = {The European Physical Journal A},
  publisher = {Springer Science and Business Media LLC},
  author = {Colò,  G. and Garg,  U. and Sagawa,  H.},
  year = {2014},
  month = feb 
}

@article{HZheng16PRC,
  title = {Dipole response in neutron-rich nuclei with new Skyrme interactions},
  author = {Zheng, H. and Burrello, S. and Colonna, M. and Baran, V.},
  journal = {Phys. Rev. C},
  volume = {94},
  issue = {1},
  pages = {014313},
  numpages = {14},
  year = {2016},
  month = {Jul},
  publisher = {American Physical Society},
  doi = {10.1103/PhysRevC.94.014313},
  url = {https://link.aps.org/doi/10.1103/PhysRevC.94.014313}
}

@article{Natsumi23PRC,
  title = {Collision integral with momentum-dependent potentials and its impact on pion production in heavy-ion collisions},
  author = {Ikeno, Natsumi and Ono, Akira},
  journal = {Phys. Rev. C},
  volume = {108},
  issue = {4},
  pages = {044601},
  numpages = {18},
  year = {2023},
  month = {Oct},
  publisher = {American Physical Society},
  doi = {10.1103/PhysRevC.108.044601},
  url = {https://link.aps.org/doi/10.1103/PhysRevC.108.044601}
}

@article{CXu10PRC,
  title = {Symmetry energy, its density slope, and neutron-proton effective mass splitting at normal density extracted from global nucleon optical potentials},
  author = {Xu, Chang and Li, Bao-An and Chen, Lie-Wen},
  journal = {Phys. Rev. C},
  volume = {82},
  issue = {5},
  pages = {054607},
  numpages = {5},
  year = {2010},
  month = {Nov},
  publisher = {American Physical Society},
  doi = {10.1103/PhysRevC.82.054607},
  url = {https://link.aps.org/doi/10.1103/PhysRevC.82.054607}
}

@article{Chabanat97NPA,
title = {A Skyrme parametrization from subnuclear to neutron star densities},
journal = {Nuclear Physics A},
volume = {627},
number = {4},
pages = {710-746},
year = {1997},
issn = {0375-9474},
doi = {https://doi.org/10.1016/S0375-9474(97)00596-4},
url = {https://www.sciencedirect.com/science/article/pii/S0375947497005964},
author = {E. Chabanat and P. Bonche and P. Haensel and J. Meyer and R. Schaeffer}
}

@article{Baran12PRC,
  title = {Pygmy dipole resonance: Collective features and symmetry energy effects},
  author = {Baran, V. and Frecus, B. and Colonna, M. and Di Toro, M.},
  journal = {Phys. Rev. C},
  volume = {85},
  issue = {5},
  pages = {051601},
  numpages = {6},
  year = {2012},
  month = {May},
  publisher = {American Physical Society},
  doi = {10.1103/PhysRevC.85.051601},
  url = {https://link.aps.org/doi/10.1103/PhysRevC.85.051601}
}

@article{ZZhang15PRC,
  title = {Electric dipole polarizability in $^{208}\mathbf{Pb}$ as a probe of the symmetry energy and neutron matter around ${\ensuremath{\rho}}_{0}/3$},
  author = {Zhang, Zhen and Chen, Lie-Wen},
  journal = {Phys. Rev. C},
  volume = {92},
  issue = {3},
  pages = {031301},
  numpages = {5},
  year = {2015},
  month = {Sep},
  publisher = {American Physical Society},
  doi = {10.1103/PhysRevC.92.031301},
  url = {https://link.aps.org/doi/10.1103/PhysRevC.92.031301}
}

@article{ZZLi21PRC,
  title = {Electric dipole polarizability in neutron-rich Sn isotopes as a probe of nuclear isovector properties},
  author = {Li, Z. Z. and Niu, Y. F. and Long, W. H.},
  journal = {Phys. Rev. C},
  volume = {103},
  issue = {6},
  pages = {064301},
  numpages = {8},
  year = {2021},
  month = {Jun},
  publisher = {American Physical Society},
  doi = {10.1103/PhysRevC.103.064301},
  url = {https://link.aps.org/doi/10.1103/PhysRevC.103.064301}
}

@article{Bennaceur05CPC,
title = {Coordinate-space solution of the Skyrme–Hartree–Fock– Bogolyubov equations within spherical symmetry. The program HFBRAD (v1.00)},
journal = {Computer Physics Communications},
volume = {168},
number = {2},
pages = {96-122},
year = {2005},
issn = {0010-4655},
doi = {https://doi.org/10.1016/j.cpc.2005.02.002},
url = {https://www.sciencedirect.com/science/article/pii/S0010465505002304},
author = {K. Bennaceur and J. Dobaczewski}
}

@misc{ZZLi_private,
  author       = {Li, Z. Z.},
  title        = {{SHFB + QRPA} calculation of the spin-orbit effect on the electric dipole polarizability},
  year         = {2026},
  note         = {Private communication}
}

\end{document}